\documentclass[aps,prd,superscriptaddress,nofootinbib,amsmath,amsfonts, preprintnumbers, groupedaddress, showpacs,
10pt,english]{revtex4-1}
\usepackage{amsmath}
\usepackage{amssymb}
\usepackage{babel}
\usepackage{wrapfig}
\usepackage{cancel}

\usepackage{relsize,exscale}
\makeatletter

\usepackage{array,multirow,graphicx}
\usepackage{dcolumn}
\usepackage{newlfont}
\usepackage{bm}
\usepackage[colorlinks,citecolor=blue,urlcolor=blue,linkcolor=blue]{hyperref}
\usepackage[figtopcap]{subfigure}
\usepackage{color}

\newcommand{\beq}{\begin{equation}}
\newcommand{\eeq}{\end{equation}}
\newcommand{\bea}{\begin{eqnarray}}
\newcommand{\eea}{\end{eqnarray}}

\newcommand{\alp}{\alpha}

\newcommand{\del}{\delta}
\newcommand{\bet}{\beta}

\newcommand{\comm}[1]{}

\usepackage{scalerel}
\usepackage{tikz}
\usetikzlibrary{svg.path}
\definecolor{orcidlogocol}{HTML}{A6CE39}
\tikzset{
  orcidlogo/.pic={
    \fill[orcidlogocol] svg{M256,128c0,70.7-57.3,128-128,128C57.3,256,0,198.7,0,128C0,57.3,57.3,0,128,0C198.7,0,256,57.3,256,128z};
    \fill[white] svg{M86.3,186.2H70.9V79.1h15.4v48.4V186.2z}
                 svg{M108.9,79.1h41.6c39.6,0,57,28.3,57,53.6c0,27.5-21.5,53.6-56.8,53.6h-41.8V79.1z M124.3,172.4h24.5c34.9,0,42.9-26.5,42.9-39.7c0-21.5-13.7-39.7-43.7-39.7h-23.7V172.4z}
                 svg{M88.7,56.8c0,5.5-4.5,10.1-10.1,10.1c-5.6,0-10.1-4.6-10.1-10.1c0-5.6,4.5-10.1,10.1-10.1C84.2,46.7,88.7,51.3,88.7,56.8z};}}
\newcommand\orcid[1]{\href{https://orcid.org/#1}{\mbox{\scalerel*{
\begin{tikzpicture}[yscale=-1,transform shape]
\pic{orcidlogo};
\end{tikzpicture}
}{|}}}}
\begin{document}
\allowdisplaybreaks[4]
\tolerance=5000

\date{\today}
\title{Structure, maximum mass, and stability of compact stars in f(Q,T) gravity}
\author{G.~G.~L.~Nashed$^{1}$~\orcid{0000-0001-5544-1119}}
\email{nashed@bue.edu.eg}
\author{Tiberiu Harko~}
\email{tiberiu.harko@aira.astro.ro}
\affiliation {$^{1)}$Centre for Theoretical Physics, The British University, P.O. Box
43, El Sherouk City, Cairo 11837, Egypt\\
$^{2)}$ Department of Theoretical Physics, National Institute of Physics
and Nuclear Engineering (IFIN-HH), Bucharest, 077125 Romania\\
$^{3)}$Department of Physics, Babes-Bolyai University,
Kogalniceanu Street, Cluj-Napoca 400084, Romania\\
$^{4)}$Astronomical Observatory, 19 Ciresilor Street, Cluj-Napoca 400487, Romania}

\begin{abstract}
Physically based changes to general relativity (GR) often predict significant differences in how spacetime behaves near massive neutron stars. One of these modifications is represented by $f(\mathcal{Q}, { \mathcal{T}})$, with $\mathcal{Q}$ being the non-metricity   and ${ \mathit{T}}$  represents  the energy-momentum tensor trace. This theory is  viewed as a neutral expansion of GR. Neutron stars weighing more than 1.8 times the mass of the Sun, when observed as radio pulsars, provide valuable opportunities to test fundamental physics under extreme conditions that are rare in the observable universe and cannot be replicated in experiments conducted on land. We derive an exact  solution  through utilizing the form $f(\mathcal{Q}, { \mathcal{T}})=\mathcal{Q}+\psi { \mathcal{T}}$, where $\psi$ represents a dimensional expression. We elucidate that all physical quantities within the star can be expressed using the dimensional parameter $\psi$ and the compactness, which is defined as $C=\frac{ 2GM}{Rc^2}$. We set $\psi$ to a maximum value of $\psi_1=\frac{\psi}{\kappa^2}=-0.04$ in the negative range, based on observational constraints related to  radius and mass of the pulsar ${\textit SAX J1748.9-2021}$.  Here, ${\mathrm \kappa^2}$ represents the coupling constant of Einstein, defined as ${\mathrm \kappa^2=\frac{8\pi G}{c^4}}$.   The solution we derived results in a stable compact object without violating the conjectured sound speed condition $c_s^2\leq\frac{c^2}3$, unlike in GR.It is crucial to mention that no equations of state were assumed in this investigation. Nevertheless, our model fits nicely with linear form. Generally, when $\psi$ is negative, the theory predicts a star with a slightly larger size compared to GR for the same mass. The difference in predicted size between the theory with a negative $\psi$ and GR for the same mass is attributed to an additional force. This force results from the interaction between matter and geometry, mitigating the influence of the gravitational force.
 We determined the highest level of compactness permitted by the strong energy condition for
 $f(\mathcal{Q}, { \mathcal{T}})$ and for GR, resulting in values of $C = 0.514$ and $0.419$, respectively. This is approximately 10\% above the forecast made by GR. Additionally, we estimated the highest mass to be approximately $4.66 M_\odot$ with a radius of about $14.9$ km, assuming a surface density at saturation nuclear density of $\rho_{\text{nuc}} = 4\times 10^{14}$ g/cm$^3$.
\end{abstract}
\date{}
\maketitle
\tableofcontents
\section{Introduction}

 Various independent cosmological observations suggest that the Universe is undergoing acceleration \cite{SupernovaSearchTeam:1998fmf, SupernovaCosmologyProject:1998vns, SupernovaCosmologyProject:2011ycw}. The accelerated expansion is explained by incorporating ($\Lambda$),  which represents the cosmological constant,  into the action of Einstein-Hilbert, allowing General Relativity to account for the Universe's recent accelerated growth.  General Relativity (GR) of Einstein has been highly successful, but it faces difficulties due to cosmological observations related to dark energy and dark matter. To address this, two primary approaches are commonly explored.  The first approach modifies the matter sector by adding extra dark energy components to the Universe's energy budget. The second approach extends the geometric framework of General Relativity by modifying the Einstein-Hilbert action. A well-known geometric extension of GR is $f(R)$ gravity, that depend on curvature \cite{Nojiri:2017ncd,Sotiriou:2008rp,DeFelice:2010aj,Harko:2011kv} .

  Other alternative method to broaden the geometric aspect is by considering torsion and non--metricity. Gravitational theories based on torsion that are equivalent to General Relativity are referred to as the Teleparallel Equivalent of General Relativity (TEGR) \cite{Unzicker:2005in,Hayashi:1979qx, Sauer:2004hj}, theories based on non-metricity are known as the Symmetric Teleparallel Equivalent of General Relativity (STEGR). \cite{Nester:1998mp, BeltranJimenez:2018vdo}. TEGR gravity is founded on torsion rather than the curvature of space-time to describe gravitational interactions.   When both the curvature and non-metricity tensors vanish, the spin connection becomes constrained, reflecting the specific geometric properties of the theory.  These constraints allow for the use of the Weitzenböck connection, where all spin connection components vanish, leaving only the tetrad components as the fundamental variables. This is considered a gauge choice in TEGR. Crucially, it does not alter the physics, as any valid choice that satisfies the teleparallelism condition will produce the same action, differing only by a surface term.

  Non-metricity, instead of curvature and torsion, is used to explain gravitational effects in STEG. In the context of teleparallelism, a coincidence gauge can be utilized, which simplifies the main variable to the metric tensor. A variation of STEG is $f(\mathcal{Q})$ gravity, which bears resemblance to $f(R)$ gravity \cite{BeltranJimenez:2017tkd,BeltranJimenez:2018vdo, Heisenberg:2018vsk}. Various aspects of $f(\mathcal{Q})$ gravity have been explored in existing literature \cite{BeltranJimenez:2019tme, Anagnostopoulos:2021ydo, Barros:2020bgg, Flathmann:2020zyj}.  Extended theories of gravity have gained importance in uncovering the astrophysical and cosmological properties of the Universe. In the broader category of $f(\mathcal{Q})$ gravity, the propagation velocity and potential polarization of gravitational waves across Minkowski spacetime were explored in \cite{Hohmann:2018wxu}. Gravitational wave polarization plays a big role in how gravity behaves strongly, as mentioned by \cite{Soudi:2018dhv}. The $f(\mathcal{Q})$ theory has been tackled in various situations, Lazkoz  et. al \cite{Lazkoz:2019sjl} studied it in terms of the universe's expansion over time, Bajardi et al. \cite{Bajardi:2020fxh} considered it in bouncing universes, and Ambrosio et al. \cite{DAmbrosio:2021zpm} explored its implications for black holes. Additionally, Khyllep et. al   \cite{Khyllep:2021pcu} delved into how it affects the growth of structures in the universe. To match what we observe, researchers like Ayuso \citet{Ayuso:2020dcu} have examined different ways to describe $f(\mathcal{Q})$ theory based on observational data.

   In recent studies, the concept of gravitational decoupling (GD) has been integrated into various modified theoretical frameworks. Similar to how it functions in standard 4D classical gravity theory, GD enables the anisotropization of initial solutions. This feature offers a way to explore the effects of anisotropic stresses within compact objects.  The minimal and extended geometric deformation methods \cite{Ovalle:2018gic,Ovalle:2017fgl} were applied to analyze various aspects of compact star modeling. References \cite{Ovalle:2018umz, Ovalle:2017wqi, Ovalle:2018ans,ATLAS:2018lkz, Ovalle:2018vmg, Ovalle:2019lbs, Contreras:2019iwm, Estrada:2019aeh, Ovalle:2020kpd, Estrada:2018vrl, Contreras:2021yxe, Maurya:2020rny, Leon:2021qvs,Contreras:2021xkf} were consulted in this research. The research revealed that the collective influence of decoupling parameters led to higher masses for neutron stars. Compact star objects exhibiting masses comparable to the GW190814 event in $f(\mathcal{Q})$ gravity were pinpointed in \cite{Maurya:2022vsn}. The metricity parameter and deformation constant play crucial roles in reliably and consistently defining models for compact stellar systems. The minimal and extended gravitational decoupling methods have effectively accommodated stellar masses exceeding 2 solar masses without the need for exotic matter distributions, as discussed in \cite{Maurya:2022vsn,  Maurya:2021zvb}. Modeling compact objects using gravitational decoupling techniques has proven successful in both classical and modified gravity theories. This approach extends to modeling phenomena such as black holes and gravitational lensing, as discussed in \cite{Ovalle:2013xla}. The study employs the extended GD method along with the complete geometric deformation (CGD) method to model strange stars within the symmetric teleparallel formalism coupled with the electromagnetic field. It investigates the impact of non--metricity and coupling constants on the maximum permissible masses and radii of self-gravitating compact stars such as PSR J1614-2230, PSR J1903+6620, Cen X-3, and LMC X-4. Furthermore, the research compares these results with the observational constraints derived from GW190814.

Within such study, we have examined the feasibility and stability of Neutron star  taking into account   Krori-Barua (KB) solution in the frame of $f(\mathcal{Q})$
theory. The present study is organized in the following manner: Section \ref{sec1} presents the fundamentals of $f(\mathcal{Q})$ theory taking into account anisotropic matter sector. Section \ref{int} provides a geometric discussion  of the metric and investigates the determination of unknown constants. In Section \ref{phy}, various physical features are analyzed to assess the viability of compact stars.  In Section \ref{Sec:EoS_MR}, we delve into the induced EoSs that govern the matter sector and discuss the mass-radius relationship. In Section \ref{con} we provide a summary of our results.

\section{$f(\mathcal{Q,T})$ \,\, Theory: Field Equations}\label{sec1}

We will now briefly examine the STEGR theory. Within this framework, we define a spacetime configuration represented by ($\mathcal{M}, g_{\mu \nu}, {\Gamma^\alpha}{\mu \nu}$), with $g_{\mu \nu}$ symbolizing a metric tensor that adheres to a distinct signature $(-1, +1, +1, +1)$, while ${\Gamma^\alpha}_{\mu \nu}$ stands for a non-specific affine connection. In this scenario, we express the torsion tensor as follows:
    \begin{eqnarray}
            && \hspace{-5.3cm} {\mathcal T^{\alpha}_{\,\,\,\mu \nu}}=2\Gamma^\alpha_{\,\,\,[\mu \nu]}=\Gamma^{\alpha}_{\,\,\,\mu \nu}-\Gamma^{\alpha}_{\,\,\,\nu \mu} \,,  \label{eq.1}
    \end{eqnarray}
where the Riemann curvature tensor is defined as,
    \begin{eqnarray} \label{eq.2}
        && \hspace{-3.7cm} {\mathcal {R^\alpha}_{\beta \mu \nu}=\partial_\mu {\Gamma^\alpha}_{\nu \beta}-\partial_\nu {\Gamma^\alpha}_{\mu \beta} + {\Gamma^\rho}_{\nu \beta} {\Gamma^\alpha}_{\mu \rho} - {\Gamma^\rho}_{\mu \beta} {\Gamma^\alpha}_{\nu \rho}}\,.
    \end{eqnarray}
We defined the non-metricity of the connection as:
    \begin{eqnarray}\label{eq.3}
        && \hspace{-3.2cm}\mathcal{Q}_{\alpha \mu \nu}\equiv\bigtriangledown_{_\alpha}g_{\mu \nu}=\partial_\mu\, g_{\nu \alpha}-\Gamma^\beta_{\,\,\,\mu \nu}\, g_{\beta \alpha}-\Gamma^\beta_{\,\,\,\mu \alpha} \,g_{\nu \beta}\,,
    \end{eqnarray}
with  $\nabla_{\mu}$ being the covariant derivative. Therefore, we defined  the affine connection as:
    \begin{eqnarray}  \label{eq.4}
        && \hspace{-6.9cm}{\Gamma^\alpha}_{\mu \nu}=\mathring{\Gamma^\alpha}_{\,\,\,\mu \nu } - {S^\alpha}_{\,\,\,\mu \nu}\,,
    \end{eqnarray}
with  $\mathring{\Gamma^\alpha}_{\,\,\,\mu \nu}$ being the Levi-Civita connection, which is defined as:
    \begin{eqnarray} \label{eq.5}
        &&\hspace{-4.2cm} \mathring{\Gamma^\alpha}_{\,\,\,\mu \nu} = \frac{1}{2}g^{\beta \alpha}\left(\partial_\mu \,g_{\alpha \nu}+\partial_\nu \,g_{\alpha \nu}-\partial_\alpha \,g_{\mu \nu}\right)\,,
    \end{eqnarray}
where the disformation $L^{\alpha}_{\mu \nu}$ takes the form:
    \begin{eqnarray} \label{eq.6}
        &&\hspace{-4cm} {L^\alpha}_{\mu \nu} \equiv -\frac{1}{2} g^{\alpha \beta} \left(\mathcal{Q}_{\mu \beta \nu}+\mathcal{Q}_{\nu \beta \mu}+\mathcal{Q}_{\beta \mu  \nu} \right)={L^\alpha}_{\nu \mu}\,.
    \end{eqnarray}
The superpotential is defined, using the non-metricity tensor, as:
    \begin{eqnarray} \label{eq.7}
        && \hspace{-1.5cm}P^{\alpha \mu \nu}=-\frac{1}{4} \mathcal{Q}^{\alpha \mu \nu} + \frac{1}{2} \mathcal{Q}^{(\mu \nu)\alpha}+\frac{1}{4}(\mathcal{Q}^\alpha-\tilde{\mathcal{Q}}^\alpha) g^{\mu \nu}-\frac{1}{4}\delta^\alpha {{}^({}^\mu}\mathcal{Q}^{\nu)}\,.
    \end{eqnarray}
The vectors $\mathcal{Q}^\alpha$ and $\tilde{\mathcal{Q}}^\alpha$ represent two distinct components extracted from  $\mathcal{Q}_{\alpha \mu \nu}$, which is given by:
    \begin{eqnarray} \label{eq.8}
        && \hspace{-5.3cm}\mathcal{Q}_{\alpha}\equiv \mathcal{Q}_{\alpha\,\,\,~\mu}^{\,\,~\mu},~~~~~~~~\; \tilde{\mathcal{Q}}^\alpha=\mathcal{Q}^{\,\,\,\,\alpha \mu}_{\mu}\,.
    \end{eqnarray}
We define the non-metricity scalar tensor as:
    \begin{eqnarray} \label{eq.9}
        && \hspace{-3.7cm}\mathcal{Q}=-g^{\mu \nu}\left(L^{\alpha}_{\beta \nu} L^{\beta}_{\mu \alpha}-L^{\beta}_{\alpha \beta} L^{\alpha}_{\mu \nu}\right)=-\mathcal{Q}_{\alpha \mu \nu} P^{\alpha \mu \nu}\,.
    \end{eqnarray}
Now we are ready to use the above ingredients tools to give a brief description of $f(\mathcal{Q,T})$ theory.

\subsection{$f(\mathcal{Q,T})$ Gravity}
The gravitational action integral in STEGR $f(\mathcal{Q},\mathcal{T})$ gravity, as described in \cite{BeltranJimenez:2017tkd,Zhao:2021zab}, is as follows:\begin{equation}
S_{f\mathcal{(Q,T)}}=\int\sqrt{-g}d^{4}x\Big[f\left(\mathcal{{Q,T}}\right)+2\kappa^2\mathcal{L}_{m}\Big]
\label{action_f(Q)}.
\end{equation}
In this study, the gravitational theory  $f(\mathcal{Q,T})$ is introduced as a general smooth function that depends on  $\mathcal{Q}$ and $\mathcal{T}$. The matter Lagrangian density is represented by $\mathcal{L}_m$. Since $\mathcal{L}_m$ is metric-dependent, the energy momentum tensor (EMT)   is expressed as

\begin{equation}
 { \mathcal{T}_{\mu\nu}}=\frac{-2}{\sqrt{-g}}\frac{\delta \left(\sqrt{-g} \mathcal{L}_m \right)}{\delta g^{\mu\nu}}\,.
\end{equation}
We model the star's interior as an anisotropic fluid, with the EMT expressed in the following form:
\begin{equation}\label{Tmn-anisotropy}
    { \mathcal{T}{^\mu}{_\nu}=  (p_{2}+\rho c^2)w{^\mu} w{_\nu}+p_{2} \delta ^\mu _\nu + (p_{1}-p_{2}) v{^\mu} v{_\nu}}\,,
\end{equation}
where ${\mathcal \rho}$ represents the density, ${ p}_{1}$ and
 ${ p}_{2}$ are the radial and tangential  components of the  pressures,  ${  w}_\alpha$ is perpendicular to the tangential pressure  and ${v}^\alpha$ represents a unit space-like vector pointing. The   EMT
 ${\mathcal{T}_{\mu\nu}}$ in this study takes the form ${ \mathcal{T}}{^\alpha}{_\beta}=diag(-\rho c^2,\,p_{1},\,p_{2},\,p_{2})$.

The gravitational field equations are constructed by varying the modified Einstein-Hilbert action (\ref{action_f(Q)}) about the metric tensor $g_{\mu \nu}$ to yields
    \begin{eqnarray} \label{eq.11}
        && \hspace{-0.75cm}\frac{2}{\sqrt{-g}}\bigtriangledown_\alpha\left(\sqrt{-g}\,f_\mathcal{Q}\,P^\alpha_{\,\,\,\,\mu \nu}\right)+\frac{1}{2}g_{\mu \nu}f+f_\mathcal{T}\left(\mathcal{T}_{\mu \nu}+\Theta_{\mu \nu}\right)
        +f_\mathcal{Q}\big(P_{\mu\alpha \beta}\,\mathcal{Q}_\nu^{\,\,\,\,\alpha \beta}-2\,\mathcal{Q}_{\alpha \beta \mu}\,{P^{\alpha \beta}}_{\nu}\big)=\kappa^2 \mathcal{T}_{\mu \nu},~~~~
    \end{eqnarray}
where $f_\mathcal{Q}$, $f_\mathcal{T}$, denote the partial derivative of $f = f (\mathcal{Q, T})$ with respect to $\mathcal{Q}$ and $\mathcal{T}$
respectively     and $\Theta_{\mu\nu}$ is defined as:
   \begin{align}
\mathit{\Theta_{\mu\nu} \equiv g^{\alp\bet}  \frac{\del{ {T}}_{\alp\bet}}{\del g^{\mu\nu}}=-2 { {T}}_{\mu\nu}+g_{\mu\nu} {L}_m-2g^{\alp\bet}\frac{\partial^2 {L}_m}{\partial g^{\mu\nu} \partial g^{\alp \bet}}}\,.
\end{align}

It is well established that for perfect fluids, there is not universally agreed definition for $\mathcal{L}_m$ \cite{Faraoni:2009rk, Harko:2011kv, Avelino:2018rsb}. Therefore, we must make an assumption to proceed. Following the approach  presented in \cite{Harko:2011kv}, we suppose  $\mathcal{L}m = p$. Given such assumption, we obtain: $\Theta{\mu\nu} = -2 { \mathcal{T}}{\mu\nu} + p g{\mu\nu}$.

The field equations referred  in Eq.~\eqref{eq.11} can be rephrased into their corresponding effective GR representation. In this representation, the Einstein tensor $G_{\mu \nu}$ is situated on the left side of the equation. The right terms comprises the efficient EMT that incorporates both the impacts of matter fields and the contributions arising from non-metricity elements.

\beq
G_{\mu\nu}=\frac{1}{f_\mathcal{Q}}\left[{\kappa^2 \mathcal{T}}_{\mu\nu} +\frac{1}{2}g_{\mu\nu}\left[f(\mathcal{Q})-f_{\mathcal{Q}}
\left(\mathcal{Q}\right)\mathcal{Q}\right]-2f_{\mathcal{Q} \mathcal{Q}}\left(\mathcal{Q}\right)P_{\phantom{\alpha}\mu\nu}^{\alpha}\partial_{\alpha}\mathcal{Q}-f_\mathcal{T}\left(\mathcal{T}_{\mu \nu}+\Theta_{\mu \nu}\right) \right] \equiv { \mathcal{T}}^{(\text{eff})}_{\mu\nu}\,.
\label{Gmn_Teff}
\eeq
In general, when $f({ \mathcal{Q, T}})$ becomes   arbitrary, the components of ${ \mathcal{T}}^{(\text{eff})}_{\mu\nu}$ are highly complex, making it challenging to derive an exact analytical solution in this framework. It is important to note, however, that for a specific model
\beq
{\textit f( \mathcal{Q,T})= \mathcal{Q}+S( \mathcal{T})}\,,
\label{f(R,T) separable}
\eeq
that is the linear form of the non-metricity and excludes any mixing terms between ${ \mathcal{Q}}$ and ${ \mathcal{T}}$, the impact ${ \mathcal{Q}}$  to the effective EMT become much simpler. As a result, the formulation of the field equations simplifies to:
\beq
G_{\mu\nu} = \kappa^2 { \mathcal{T}}_{\mu\nu}+\frac{S}{2}g_{\mu\nu}+S_{ \mathcal{T}} \left({ \mathcal{T}}_{\mu\nu}-p\,g_{\mu\nu} \right)\,.
\label{GR+T effects EOM}
\eeq
Now we assume, $f({ \mathcal{Q,T}})$ to be:
\beq
f({ \mathcal{Q,T}})=\mathcal{Q}+\psi{ \mathcal{T}}\,,
\label{eq:linear_f(R,T)}
\eeq
where $\psi$ is which is a dimensional constant parameter. In this linear case, an analytical solution for the field equations can be readily derived  (\cite[see also][]{Hansraj:2018jzb,Bhar:2021uqr,Pretel:2020oae,Pretel:2021kgl}).

\section{Interior spherically symmetric solution}\label{int}
Compact stars are intriguing entities that form as a result of the gravitational collapse of large stars. Grasping their structure, composition, and behavior is essential for progressing in our understanding of basic physics and astrophysics.
 Krori-Barua (KB) solutions offer understanding  the way in which material acts in extreme situations inside compact stars. These suggested ideas that can anticipate unique gravitational wave patterns that can be compared to actual observations to confirm or improve the models.

 Now we employ the following line element representing a static spherically symmetric geometry::
\beq
ds^2=-e^{\zeta(r)}dt^2+e^{\zeta_1(r)}dr^2+r^2d\theta^2+r^2\sin^2\theta d\phi^2\,,
\label{line element}
\eeq
where $\zeta(r)$ and $\zeta_1(r)$ are functions of the radial coordinate.

For the metric equation (\ref{line element}), we can calculate the non-metricity scalar as follows:
    \begin{eqnarray} \label{eq.26}
        &&\hspace{-5.6cm} \mathcal{Q}=\frac{(e^{\zeta(r)}-1) \left(\zeta'+\zeta'_1\right)}{r},
    \end{eqnarray}
In the expression of $\mathcal{Q}$,  $'$ represents the derivative w.r.t $r$, and $\mathcal{Q}$ is based on non-zero affine connections\footnote{It is crucial to note that if the line element given by Eq. (\ref{line element}) is flat, i.e., $\zeta=\zeta_1=0$ then the non-metricity given by Eq. (\ref{eq.26}) is vanishing which is necessary  for physical STEGR theory }. Based on the equations of motion (\ref{eq.11}) for the anisotropic fluid (\ref{Tmn-anisotropy}), the independent components are as follows:
\begin{eqnarray}
  &&{\kappa}^{2} \rho  =\frac{5\left\{ 2\,\psi_1\,{r}^{2}\zeta''  +\psi_1\,{r }^{2} \zeta'^{2}-r\psi_1\, \left( \zeta'_1r -4\right) \zeta' +{\frac {32}{5}}\, \left( \frac{3}{16}+\psi_1 \right)[ \zeta'_1 r+{e^{\zeta_1 }}-1 ]\right\}}{\,{e^{\zeta_1}} \left( 96\,{\psi_1}^{2}{r}^{2}+60\,\psi_1\,{r}^{2}+6\,{r}^{2}\right)}
 ,\label{19}\nonumber\\
 && {\kappa}^{2} p_1=-\frac{5 \left\{ 2\,\psi_1\,{r}^{2}\zeta'' +\psi_1\,{r}^{2} \zeta'^{2}-r \left( {\frac {28}{5}}\,\psi_1
+\psi_1\, \zeta'_1r+\frac{6}5\,\right) \zeta' -{\frac {16}{5}}
\,\psi_1\, \zeta'_1 r+{\frac
{32}{5}}\, \left( \frac{3}{16}\,+\psi_1 \right)  \left( {e^{\zeta_1 }}-1 \right)  \right\}}{{e^{\zeta_1}} \left( 96\,{\psi_1}^{2}{r}^{
2}+60\,\psi_1\,{r}^{2}+6\,{r}^{2} \right)}\,,\nonumber\\
&&{\kappa}^{2} p_2=\frac{14\, \left\{ 2\,{r}^{2} \left( \psi_1+\frac{3}{14} \right)\zeta'' +{r}^{2} \left( \psi_1+\frac{3}{14}  \right)  \zeta'^{2}-r \left( r \left( \psi_1+\frac{3}{14} \right) \zeta'_1 -\frac{4\psi_1}7-\frac{3}7 \right) \zeta' -{\frac {8r}{7}} \left( \frac{3}8+\alpha \right) \zeta'_1 +{ \frac {16}{7}}\,\psi_1\, \left( {e^{\zeta'_1 }}-1 \right)  \right\}}{ {e^{\zeta_1 }} \left( 192\,{\psi_1 }^{2}+120\,\psi_1+12 \right) {r}^2}\,,\nonumber\\
&&
\end{eqnarray}
where  $\psi_1=\frac{\psi}{\kappa^2}$. Therefore, we derive the anisotropy parameter, $\mathit{\Delta(r) = p_2-p_1}$, as:
\begin{equation}\label{eq:Delta1v}
\mathrm{\Delta(r) =\frac {2\zeta'' {r}^{2}+\zeta'^2{r}^{2}- \left[{\zeta'_1}{r}+2 \right] \,r\zeta'+4 e^{\zeta_1}-2{\zeta_1}r-4}{4\kappa^2\left(2\psi_1+1 \right){r}^{2}e^{\zeta_1}}}.
\end{equation}
Outstandingly, the matter-geometry coupling due to the component of the trace is not share to $\Delta$ given by Eq. \eqref{eq:Delta1v} just the spherically symmetric spacetime configuration is supposed \cite{Nashed:2022zyi}. Nevertheless, when $\psi_1 \neq 0$  a little change in anisotropy is expected. Therefore, shifts from GR because of the matter-geometry coupling cannot be spoiled with different anisotropic effects.  When the dimensionless $\psi_1$ vanishes, the differential equations \eqref{19} will be identical with Einstein GR equation of motions for an interior spherically symmetric  spacetime \cite[c.f.,][]{Roupas:2020mvs}.

 Equations (\ref{19}) comprises three independent  differential equations having  5 unknowns $\mathrm{\zeta}$, $\mathrm{\zeta_1}$, $\mathrm{\rho}$, $\mathrm {p_1}$ and $\mathrm{p_2}$. Thus, we require two   constraints  to close Eq.(\ref{19}).Recently, the study of stellar objects using KB solutions has gained popularity because of their unique behavior.  Thus, we present the Krori-Barua (KB)  solution given as \cite{Barua:1975dxq}:
\begin{equation}\label{eq:KB}
    \zeta(r)=\frac{\varepsilon\, r^2}{R^2}+\varepsilon_1 \equiv \varepsilon \,x^2+\varepsilon_1,\,  \qquad \qquad \qquad {\zeta_1}(r)=\frac{\varepsilon_2\, r^2}{R^2}\equiv \varepsilon_2\, x^2,
\end{equation}
where the radius without units $x$ is specified as \[0 \leq \frac{r}{R}=x \leq 1\,,\]  with $R$ being  the   stellar's radius. Moreover,  \{$\varepsilon, \varepsilon_1, \varepsilon_2$\} are  fixed by matching conditions on the boundary surface of the star. Moreover, we define the dimensionless variables
\begin{equation}
    \widetilde{\rho}(r)=\frac{\rho(r)}{\rho_{\ast}}\,, \qquad \widetilde{p}_1(r)=\frac{p_1(r)}{\rho_{\ast} c^2}\,, \qquad \widetilde{p}_2(r)=\frac{p_2(r)}{\rho_{\ast} c^2}\,, \qquad \widetilde{\Delta}(r)=\frac{\Delta(r)}{\rho_{\ast} c^2}\,,
\end{equation}
where $\rho_{\ast}$ is the characteristic density defined as:
\begin{equation}
    \rho_{\ast}=\frac{1}{\kappa^2 c^2 R^2(1+10\psi_1+16\psi_1{}^2)}\,.
\end{equation}
Thus the field equations \eqref{19} read
\begin{eqnarray}
\widetilde{\rho}&=& \frac{e^{-\varepsilon_2 x^2}}{x^2}(e^{\varepsilon_2 x^2}-1+2\varepsilon_2 x^2) \nonumber\\
&+&\frac{2\psi_1}{3x^2}\left[\left(5\varepsilon(\varepsilon-\varepsilon_2)x^4 +(16\varepsilon_2+15\varepsilon)x^2 -8\right)e^{-\varepsilon_2 x^2} +8\right],\nonumber\\[8pt]
\widetilde{p}_1&=&\frac{e^{-\varepsilon_2 x^2}}{x^2}(1-e^{\varepsilon_2 x^2}+2\varepsilon x^2) \nonumber\\
&-&\frac{2\psi_1}{3x^2}\left[\left(5\varepsilon (\varepsilon-\varepsilon_2) x^4 - (9 \varepsilon_2-8 \varepsilon) x^2 -8\right)e^{-\varepsilon_2 x^2} +8\right],\nonumber\\[8pt]
\widetilde{p}_2&=& e^{-\varepsilon_2 x^2}(2 \varepsilon-\varepsilon_2 +\varepsilon (\varepsilon - \varepsilon_2) x^2) \nonumber\\
&-&\frac{2\psi_1}{3x^2}\left[\left(7\varepsilon (\varepsilon-\varepsilon_2) x^4 - (9 \varepsilon_2-4 \varepsilon) x^2 +8\right)e^{-\varepsilon_2 x^2} - 8\right],\nonumber \\
\label{eq:Feqs2}
\end{eqnarray}
and the anisotropy factor \eqref{eq:Delta1v} becomes
\begin{equation}\label{eq:Delta2}
   \widetilde{\Delta}=\frac{e^{-\varepsilon_2 x^2}}{x^2}(1+8\psi_1)\left\{e^{\varepsilon_2 x^2}(1+\varepsilon_2 x^2)[(\varepsilon-\varepsilon_2)x^2-1]+1 \right\}.
\end{equation}
The mass content within a radius $r$ is given by:
\begin{equation}\label{11}
    \mathcal{M}(r)=4 \pi \int_0^r \rho(\widetilde{r}) \, \widetilde{r}{^2} d\widetilde r.
\end{equation}
By substituting  $\widetilde{\rho}$ given by Eq.~\eqref{eq:Feqs2} in  \eqref{11} we obtain:
\begin{equation}\label{eq:Mass}
    \mathcal{M}(x)=\frac{M}{C(1+10\psi_1+16\psi_1{}^2)}e^{-\varepsilon_2 x^2}\left[x(e^{\varepsilon_2 x^2}-1)+\psi_1 \, \varphi(x)\right].
\end{equation}
where  $C$ is the compactness parameter and the function $\varphi(x)$ are defined as:
\begin{eqnarray}\label{COMP}
C&=&\frac{2GM}{c^2 R}, \nonumber\\
\varphi(x)&=& \frac{\varepsilon_2{}^{-\frac{9}{2}}}{3}\left\{\left(16 x \varepsilon_2{}^\frac{9}{2}+\frac{15}{4} \varepsilon \sqrt{\pi} \varepsilon_2{}^2 (\varepsilon_2+\varepsilon) \text{erf}(\sqrt{\varepsilon_2} x) \right) e^{\varepsilon_2 x^2}\right.\nonumber\\
&-&\left. x \left[\left(\frac{15}{2} \varepsilon \varepsilon_{}^\frac{5}{2}+5\varepsilon_2{}^\frac{7}{2} (\varepsilon x^2 +\frac{3}{2})\right)\varepsilon\varepsilon_2{}^\frac{5}{2} -(5\varepsilon x^2-16) \varepsilon_2{}^\frac{9}{2}\right]\right\}\,.\nonumber\\
\end{eqnarray}
The function  $erf$ is defined as:
\[ erf(y)=\frac{2}{\sqrt{\pi}}\int_{y=0}^{y=t} e^{-y^2} dy\,.\]
Equations \eqref{eq:Mass} reduces to the model of GR  when $\psi_1=0$ \cite{Roupas:2020mvs,ElHanafy:2022kjl,Nashed:2023uvk}.

It  is familiar that   exterior solution  of GR and $f({ \mathcal{Q, T}})=\mathcal{Q}+\psi_1 { \mathcal{T}}$ are equivalent is given by:
\begin{equation}
    ds^2=-\left(1-\frac{2GM}{c^2r}\right) c^2 dt^2+\frac{dr^2}{\left(1-\frac{2GM}{c^2 r}\right)}+r^2 d\theta^2+r^2\,\sin^2 \theta d\phi^2.
\end{equation}
Recalling  Eq.~\eqref{eq:Feqs2},  we take:
\begin{equation}\label{eq:bo}
    \zeta(x=1)=\ln(1-C),\, \qquad \zeta_1(x=1)=-\ln(1-C).
\end{equation}
Moreover, we suppose that  $p_1=0$,  i.e.,
\begin{equation}
    \widetilde{p}_1(x=1)=0.
\end{equation}
By the use of  KB ansatz \eqref{eq:KB} and the radial pressure \eqref{eq:Feqs2} along with the above the boundary conditions, we get:
\begin{eqnarray}
\varepsilon&=&\frac{-1}{10\mathcal{C}{\psi_1}} \left( 5\mathcal{C} \psi_1\,\ln  \mathcal{C}- 9\mathcal{C} \psi_1-3\mathcal{C}\pm\sqrt{ \mathcal{C} \left[ {2\psi_1}\mathcal{C}  \ln \mathcal{C}  \left[25\psi_1-5  \left( 25 \psi_1+3 \right)\right]- \left( 241\,C-81 \right) {\psi_1}^{2}- \left( 84C-54 \right) \psi_1+9\mathcal{C}\right] } \right),\nonumber\\
\varepsilon_1&=&\frac{1}{10 \mathcal{C}{\psi_1}} \left\{3\,
 \left( 5\psi_1\,\ln\mathcal{C} -1-3\,\psi_1
 \right)  \mathcal{C}\mp\sqrt{ \mathcal{C} \left[ {2\psi_1}\mathcal{C}  \ln \mathcal{C}  \left[25\psi_1-5  \left( 25 \psi_1+3 \right)\right]- \left( 241\,C-81 \right) {\psi_1}^{2}- \left( 84C-54 \right) \psi_1+9\mathcal{C}\right] } \right\}\,,\nonumber\\
\varepsilon_2&=&-\ln\mathcal{C}\,, \qquad\mbox{where} \qquad\mathcal{C}= (1-C)\,.\label{eq:const}
\end{eqnarray}

The limit $\psi_1 \to 0$, makes  the set of constants \eqref{eq:const} to coincide with the GR version \cite{Roupas:2020mvs}
\begin{equation}
    \varepsilon=\frac{C}{2\mathcal{C}}\,, \qquad \varepsilon_1=\ln\mathcal{C}-\frac{C}{2\mathcal{C}}\,, \qquad \varepsilon_2=-\ln\mathcal{C}.
\end{equation}
Interestingly all the physical quantities of KB spacetime in a given star, $0\leq x\leq 1$, can be rewritten as dimensionless terms as  $\psi_1$ and  $C$, i.e. $\widetilde{\rho}(\psi_1,C)$, $\widetilde{p}_r(\psi_1,C)$ and $\widetilde{p}_t(\psi_1,C)$.
Moreover, it yields a chance to put a higher constraint  of the allowed $C$ for  NS  and thus  the maximum mass. This will be discussed in details in Sec. \ref{Sec:EoS_MR}.

\subsection{Radial derivatives}
Here we calculate the  derivatives of $\widetilde{\rho}$ and the comports of the pressures and get:
\begin{align}\label{eq:dens_grad}
&\widetilde{\rho}'=-\frac{2}{3{x}^{3}}\left\{40\,{e^{-\varepsilon_2\,{x}^{2}}}{x}^{4}\psi_1\,\varepsilon\,\varepsilon_2-10\,{e^{-\varepsilon_2\,{x}^{2}}}{x}^{4}\psi_1\,{\varepsilon}^{2}-16\,
\psi_1\,{e^{-\varepsilon_2\,{x}^{2}}}\varepsilon_2\,{x}^{2}-3\,{e^{-\varepsilon_2\,{x}^{2}}}\varepsilon_2\,{x}^{2}-10 \,{e^{-\varepsilon_2\,{x}^{2}}}{x}^{6}\psi_1\,\varepsilon\,{\varepsilon_2}^{2}\right.\nonumber\\
&\left.+10\,{e^{-\varepsilon_2{x}^{2}}}{x}^{6}\psi_1\,
\varepsilon_2\,{\varepsilon}^{2}+32\,{e^{-\varepsilon_2\,{x}^{2}}}{x}^{4}\psi_1\,{\varepsilon_2}^{2}+6\,{x}^{4}{e^{-\varepsilon_2\,{x}^{2}}}
{\varepsilon_2}^{2}-16\,\psi_1\,{e^{-\varepsilon_2\,{x}^{2}}}-3\,{e^{-\varepsilon_2\,{x}^{2}}}+3+16\,\psi_1\right\}\,,\nonumber\\
\end{align}
\begin{align}\label{eq:pr_grad}
  & \widetilde{p}'_1=\frac{2}{3{x}^{3}} \left\{-8\,{e^{-\varepsilon_2\,{x}^{2}}}{x}^{4}\psi_1\,\varepsilon\,\varepsilon_2-10\,{e^{-\varepsilon_2\,{x}^{2}}}{x}^{4}\psi_1\,{\varepsilon}^{2}-10\,{e^{-\varepsilon_2\,{x}^{2}}}{x}^{6}\psi_1\,\varepsilon\,
{\varepsilon_2}^{2}+10\,{e^{-\varepsilon_2\,{x}^{2}}}{x}^{6}\psi_1\,\varepsilon_2\,{\varepsilon}^{2}-16\,{e^{-\varepsilon_2\,{x}^{2} }}{x}^{4}\psi_1\,{\varepsilon_2}^{2}\right.\nonumber\\
&\left.-16\,\psi_1\,{e^{-\varepsilon_2\,{x}^{2}}}-16\,\psi_1\,{e^{-\varepsilon_2\,{x}^{2}}}\varepsilon_2\,{x}^{2}+
16\,\psi_1-3\,{e^{-\varepsilon_2\,{x}^{2}}}\varepsilon_2\,{x}^{2}-6\,
{e^{-\varepsilon_2\,{x}^{2}}}\varepsilon\,{x}^{4}\varepsilon_2-3\,{e^{-\varepsilon_2\,{x}^{2}}}+3\right\}
\,,
\end{align}
\begin{align}\label{eq:pt_grad}
&\widetilde{p}'_2=-\frac{2}{3{x}^{3}} \left\{32{e^{-\varepsilon_2{x}^{2}}}{x}^{4}\psi_1\varepsilon\varepsilon_2+9{e^{-\varepsilon_2{x}^{2}}}\varepsilon{x}^{4}\varepsilon_2-
14{e^{-\varepsilon_2{x}^{2}}}{x}^{4}\psi_1{\varepsilon}^{2}-3{e^{-\varepsilon_2{x}^{2}}}{\varepsilon}^{2}{x}^{4}-14{ e^{-\varepsilon_2{x}^{2}}}{x}^{6}\psi_1\varepsilon{\varepsilon_2} ^{2}-3{x}^{6}{e^{-\varepsilon_2{x}^{2}}}\varepsilon{\varepsilon_2}^{2}\right.\nonumber\\
&\left.+ 14{e^{-\varepsilon_2{x}^{2}}}{x}^{6}\psi_1\varepsilon_2{\varepsilon}^{2}+3{x}^{6}{e^{-\varepsilon_2{x}^{2}}}\varepsilon_2{\varepsilon}^{2}-8
{e^{-\varepsilon_2{x}^{2}}}{x}^{4}\psi_1{\varepsilon_2}^{2}-3{x}^{4}{e^{-\varepsilon_2{x}^{2}}}{\varepsilon_2}^{2}-8\psi_1
{e^{-\varepsilon_2{x}^{2}}}\varepsilon_2{x}^{2}-8\psi_1{e^{-\varepsilon_2{x}^{2}}}+8\psi_1\right\}\,,\nonumber\\
\end{align}
where $'\equiv \frac{d}{dx}$. Equations~\eqref{eq:dens_grad}, \eqref{eq:pr_grad} and \eqref{eq:pt_grad} represent the derivatives of $\widetilde{\rho}$, $\widetilde{p}_1$ and $\widetilde{p}_2$ of the EMT that are important to show  the physical stability of the compact object.

\section{Physical properties}\label{phy}

Now we are going to explore  the properties of pulsars using graphical analysis within the framework of $f(\mathcal{Q,T})$ theory.
For this purpose, we will use  the pulsar {\textit SAX J1748.9-2021}, whose  \textit{$M=1.81\pm 0.3 M_\odot$} and  \textit {$ R=11.7\pm 1.7$ km} \cite{Nashed:2023uvk,Nashed:2021pkc,Legred:2021hdx} where the solar mass $M_\odot=1.9891\times 10^{30}$ kg. Any physical stellar model must meet the following conditions labeled  (\textbf{$a_1$}) to (\textbf{$a_9$}) as:
\subsection{Matter components}\label{Sec:matt}
\begin{figure*}
\centering
\subfigure[~Density]{\label{fig:density}\includegraphics[scale=0.25]{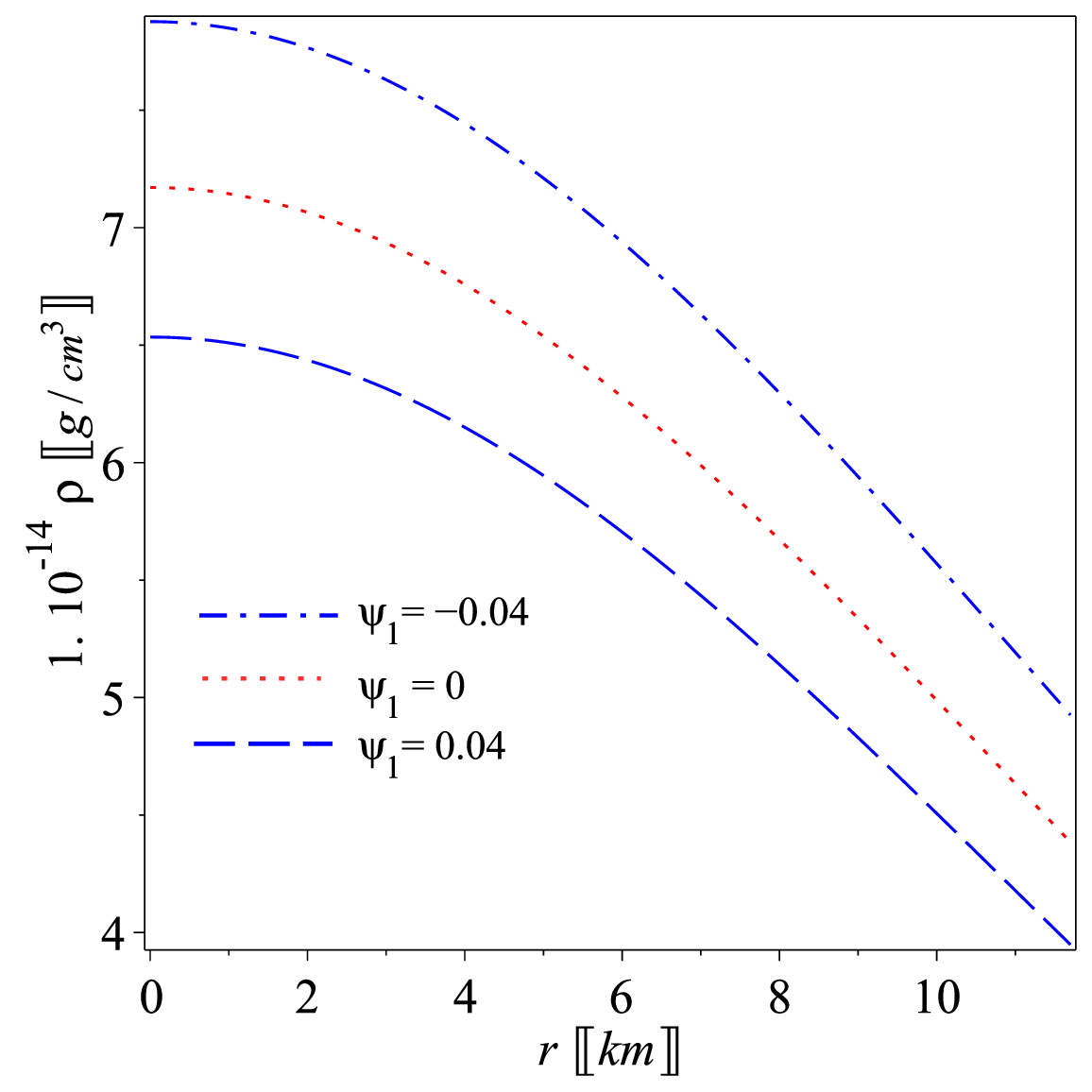}}
\subfigure[~$p_1$]{\label{fig:radpressure}\includegraphics[scale=0.25]{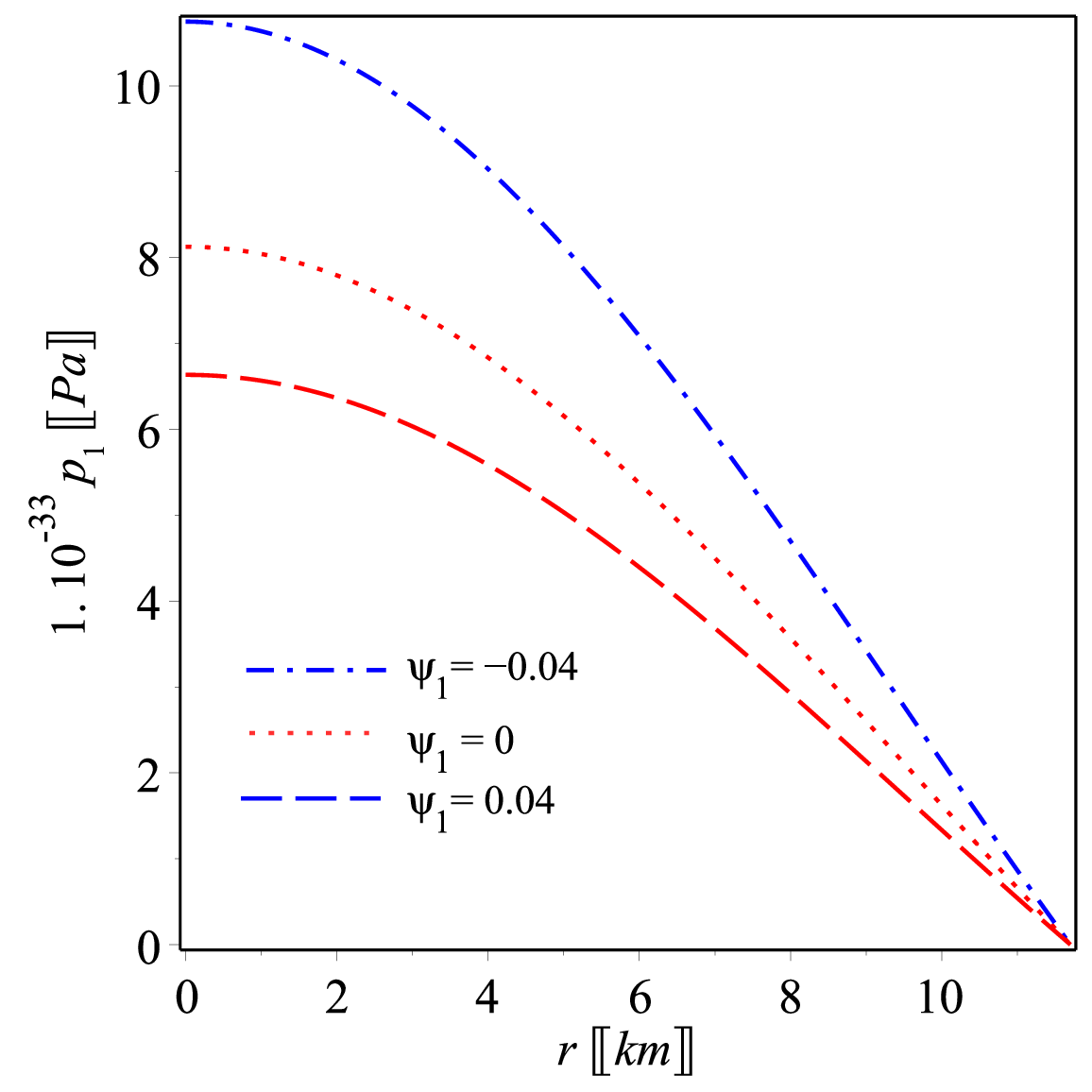}}
\subfigure[~$p_2$]{\label{fig:tangpressure}\includegraphics[scale=0.25]{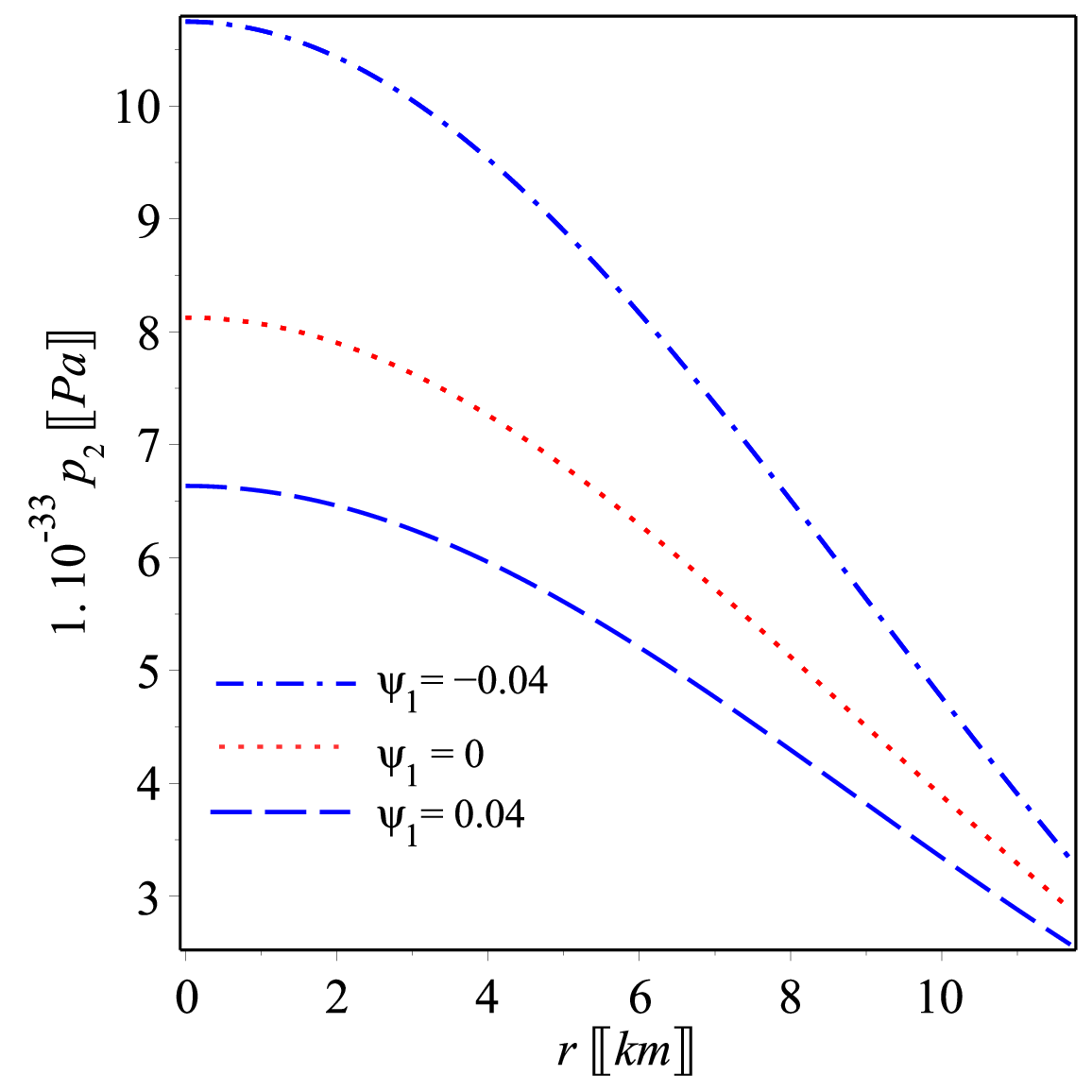}}\\
\subfigure[~$\Delta$ of  $f(\mathcal{Q,T})$]{\label{fig:anisotf}\includegraphics[scale=0.25]{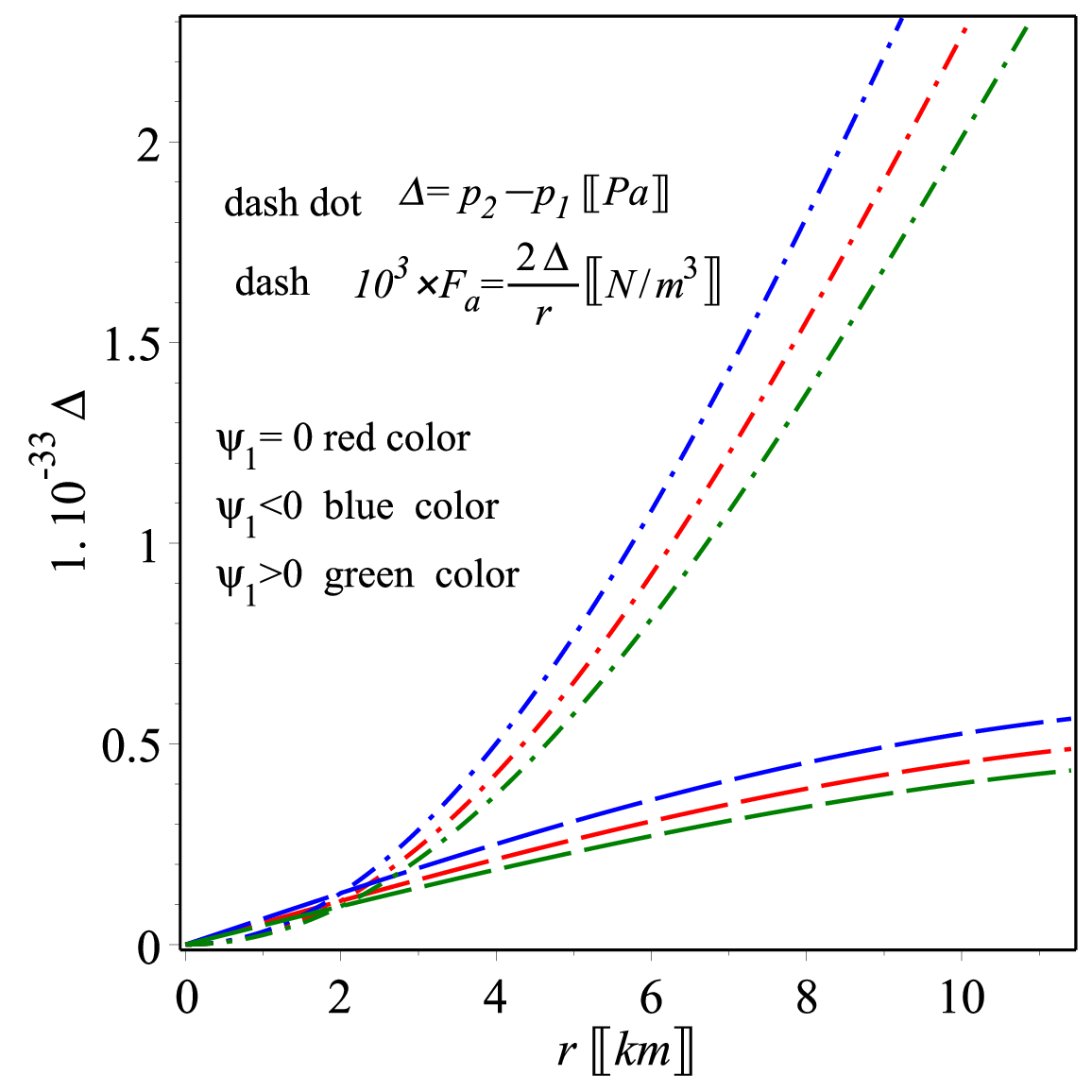}}
\subfigure[~The gradients $f(\mathcal{Q,T})$]{\label{fig:GRgrad}\includegraphics[scale=0.25]{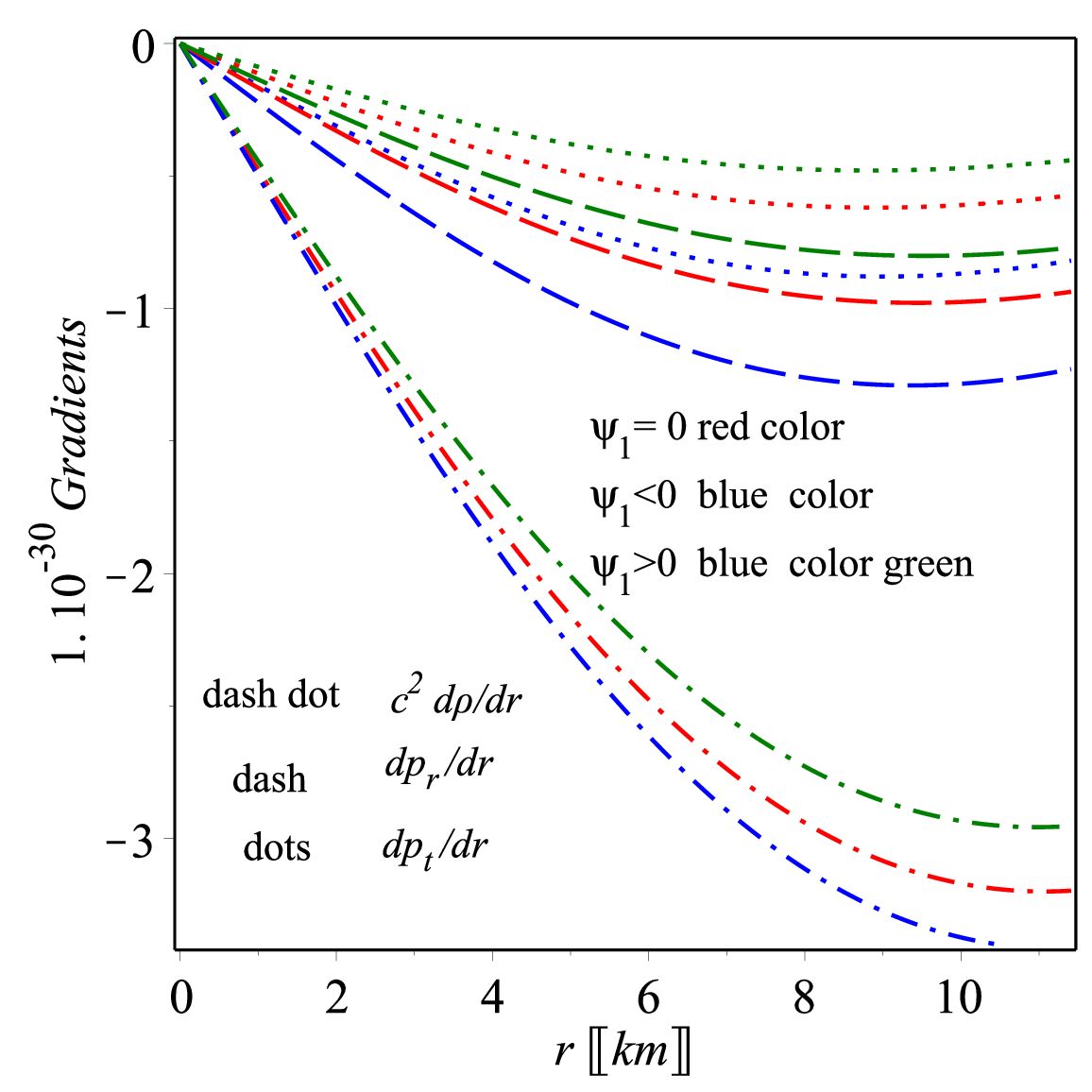}}
\caption{Patterns of  density, components of  pressures given by Eqs. \eqref{19}  of the pulsar ${\textit SAX J1748.9-2021}$ are plotted in  Figs. \subref{fig:density}--\subref{fig:tangpressure}. Figure \subref{fig:anisotf} describes  the pattern of $\Delta$ for  $f({ \mathcal{Q,T}})$ and GR  cases.  The radial gradients of these quantities are plotted in Figs. \subref{fig:GRgrad} for $f({ \mathcal{Q,T}})$ and GR scenarios . The shapes  in Fig. \ref{Fig:dens_press} confirm that the limits  listed from  (\textbf{$a_1$}) to (\textbf{$a_4$}) are verified.}
\label{Fig:dens_press}
\end{figure*}
\noindent Restriction (\textbf{$a_1$}): The EMT, $\rho$ and pressure component must possess positive pattern, specifically  ${\mathrm \rho(0 < r < R)>0}$, ${\mathrm  p_1(0 < r < R)>}0$ and ${\mathrm  p_2(0 < r < R)>0}$.

\noindent  Restriction (\textbf{$a_2$}):  The inner  solutions should be  regular. Thus,  $\rho$, $p_1$ and $p_2$   should be non-singular at the star's core, i.e.,
\begin{itemize}

  \item[I-] The energy-density is positive  as $r\to 0$, its first derivative is vanishing as $r\to 0$, its second derivative is negative as $r\to 0$, and finally its first derivative is negative  as $0<r\leq R$  \,,
  \item[II-] The radial pressure is positive  as $r\to 0$, its first derivative is vanishing as $r\to 0$, its second derivative is negative as $r\to 0$, and finally its first derivative is negative as $0<r\leq R$  \,,
  \item[III-] The tangential pressure is positive  as $r\to 0$, its first derivative is vanishing as $r\to 0$, its second derivative is negative as $r\to 0$, and finally its first derivative is negative as as $0<r\leq R$  \,.
\end{itemize}

\noindent Constrain (\textbf{$a_3$}): The radial pressure  must be vanishing at the surafce.  On the other hand, $p_2$  does not necessary to vanish at the surface. Conditions (\textbf{$a_1$})--(\textbf{$a_3$}) \textrm{are satisfied through the pulsar ${\textit SAX J1748.9-2021}$ as shown  in Fig. \ref{Fig:dens_press}\subref{fig:density}--\subref{fig:GRgrad}}.

As \ref{Fig:dens_press} \subref{fig:density},  shows that our model  approximate thee core density of a neutron star NS as  ${\mathrm \rho_\text{core}\approx 7.9\times 10^{14}}$ g/cm$^{3} \approx 4.41 \rho_\text{nuc}$ for  ${\textit SAX J1748.9-2021}$ for negative $\psi$ and ${\mathrm \rho_\text{core}\approx 6.57\times 10^{14}}$ g/cm$^{3} \approx 3.67 \rho_\text{nuc}$ for positive $\psi$ .  This means the present solution   does not rule out the probability that the pulsar at the center contains neutrons. 

\noindent Constrain (\textbf{$a_4$}): The anisotropy parameter,  $\Delta$,   must vanish at the core. Furthermore,  $\Delta$  should  have   increasing numerical value at the boundary.

 As displayed in Figs. \ref{Fig:dens_press} \subref{fig:anisotf}, condition  (\textbf{$a_4$}) $\textrm{is satisfied  for  SAX J1748.9-2021}$.
\subsection{ condition of Zeldovich}
\noindent  Condition (\textbf{$a_5$}): Based on  the  discussion presented in \cite{1971reas.book.....Z}: At the center  of the pulsar $p_1$ should be smaller than or equal to $\rho$ at the center, i.e.,
\begin{equation}\label{eq:Zel}
  \mathrm  {\frac{\widetilde{p}_1(0)}{\widetilde{\rho}(0)}\leq 1.}
\end{equation}
We derive the energy-density,  pressure components at the center like:
\begin{align}
 &\mathrm {{\widetilde\rho}(x=0) =3\varepsilon_2+2\zeta_1(8\varepsilon_2+5\varepsilon)}\,, \nonumber \\
 & \mathrm {\widetilde p_1(x=0) = {\widetilde p}_t(x=0)={2\varepsilon(1+3\varepsilon_1)-\varepsilon_2}}\,.\end{align}
For the pulsar  ${\textit SAX J1748.9-2021}$ we evaluate its compactness parameter to be $C =0.471+\pm 0.0297$. Therefore, by using Zeldovich condition \eqref{eq:Zel} we can select a real   interval of the dimensionless parameter $\psi_1$ to be $0\leq \psi_1 \leq 0.04$ for which  the parameter $\psi_1$ is expected to  coincide with GR when $\psi_1=0$.

\subsection{Mass and radius of  ${\textit SAX J1748.9-2021}$ }\label{data}
\begin{figure}
\centering
\subfigure[~The Mass function]{\label{Fig:Mass}\includegraphics[scale=0.3]{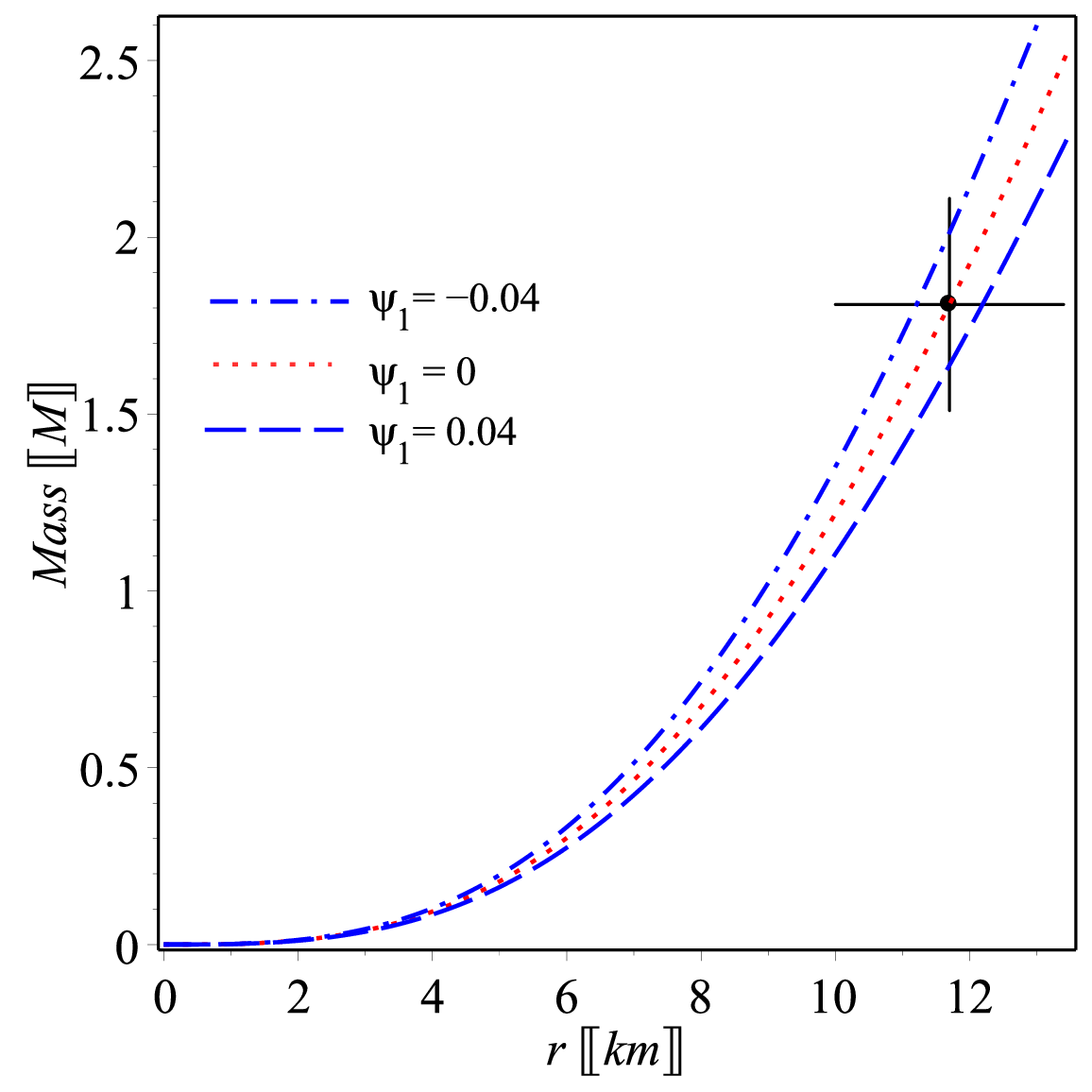}}
\subfigure[~The Compactness parameter]{\label{Fig:Comp}\includegraphics[scale=0.3]{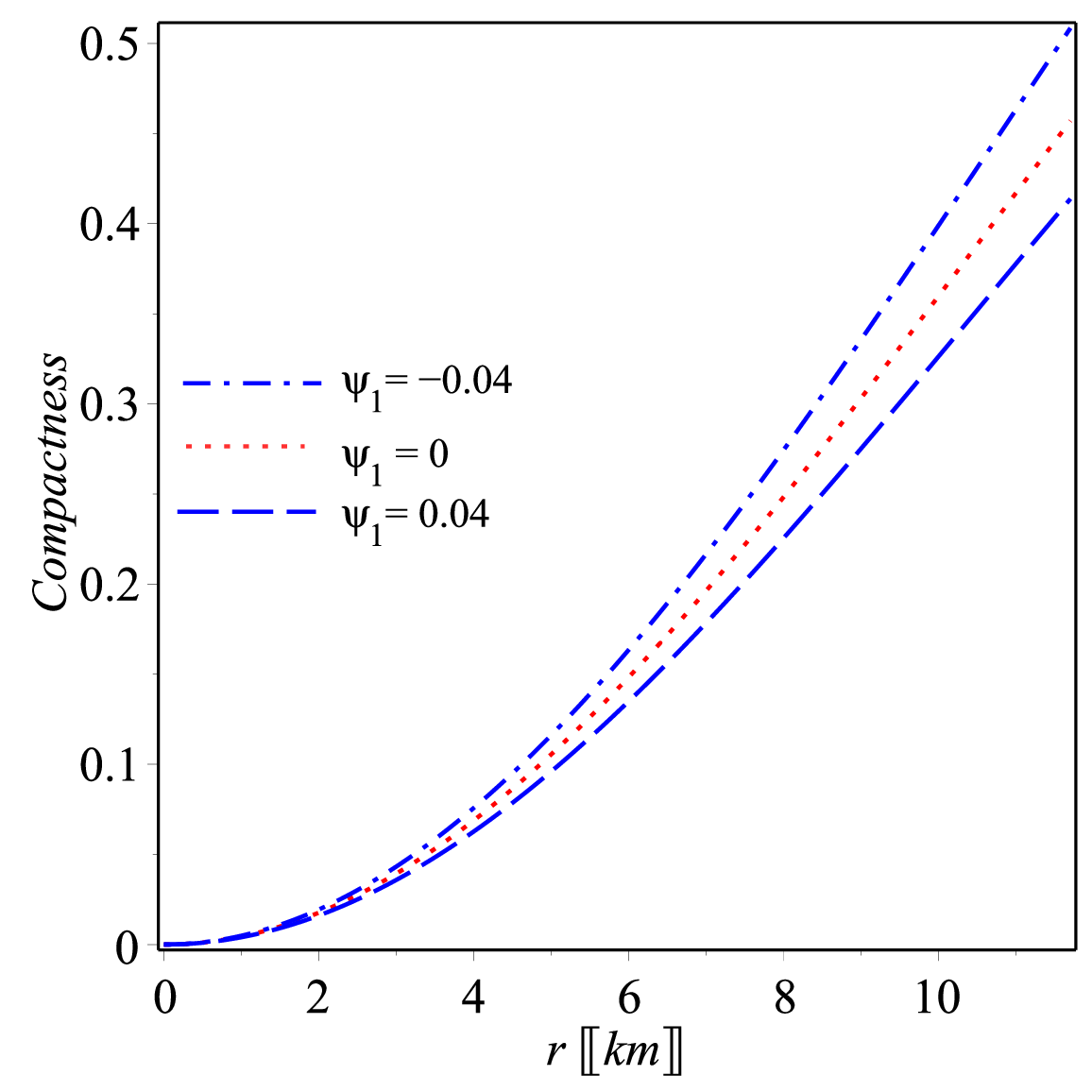}}
\caption{The figures of the mass function given by Eq.~\eqref{eq:Mass} and compactness parameter  \eqref{COMP} for the pulsar  \textit{$SAX J1748.9-2021$}. The plots are consistent  with the observational of (\textit{$ M=1.81\pm 0.3 M_\odot$} and \textit{$ R=11.7\pm1.7$ km}) \cite{Legred:2021hdx}.  When  $f({ \mathcal{Q,T}})=\mathcal{Q}+\psi_1\,{ \mathcal{T}}$   we use \{$\psi_1=0.04$, $\kappa=2.302\times 10^{-43}\,N^{-1}$, $\varepsilon =0.39$, $\varepsilon_1 =-1$, $\varepsilon_2 =0.61$\}, \{$\psi_1=-0.04$, $\kappa=2.302\times 10^{-43}\,N^{-1}$, $\varepsilon =0.45$, $\varepsilon_1 =-1.07$, $\varepsilon_2 =0.61$\} and for the case of GR  we put  ($\zeta_1=0$) and their corresponding values are \{ $\varepsilon =0.421$, $\varepsilon_1 =-1.03$,$\varepsilon_2 =0.61$\}.}
\label{Fig:Mass1}
\end{figure}

When we choose the  parameter $\psi_1=0.04$ we got   $M=1.62 M_\odot$ with corresponding $R =12.26$ km and    $C\approx 0.42$ in agreement  with {\textit ($M=1.81 \pm 0.3 M_\odot$ and $R=11.7\pm1.7$ km)} \cite{Legred:2021hdx}. This determines  the  constants presented in Eq.~(\ref{eq:const})  \{ $\varepsilon \approx 0.39$, $\varepsilon_1\approx -1$, $\varepsilon_2\approx 0.611$, $\kappa=2.302\times 10^{-43}\,N^{-1}$\}. Such numerical values satisfy   constraint \eqref{eq:Zel}. We plot the behavior of the compactness and mass  in Fig. \ref{Fig:Mass1}  \subref{Fig:Comp} and \subref{Fig:Mass}  to show  the consistent of the predicted mass-radius of the pulsar ${\textit SAX J1748.9-2021}$ and their observed values.
Therefore, $f({ \mathcal{Q,T}})={ \mathcal{Q}}+\psi_1\,{ \mathcal{T}}$ expects compactness value larger than GR for a given mass. This indicates the capability of $f({ \mathcal{Q,T}})=\mathcal{Q}+\psi_1\,{ \mathcal{T}}$ to permit  more masses or equivalently higher values of compactness while the stability conditions are verified.

\subsection{Sector of geometric}\label{Sec:geom}
\noindent  Condition (\textbf{$a_6$}): The ansatzs ${\textit g_{rr}}$ and ${\textit g_{tt}}$  do not have   singularities in the inner regime of the star $0\leq r\leq R$. The metric \eqref{eq:KB} satisfies these conditions because at the center, ${\mathrm g_{tt}(x=0)=e^{\varepsilon_1}, }$ and ${\mathrm g_{rr}(x=0)=1}$, and they are finite  in the interior  of the star $0 \leq r\leq R$.

\noindent Constrain (\textbf{$a_7$}): The   exterior and  interior  solution should joint at the boundary of pulsar.
\subsection{Constraints of energy}\label{Sec:Energy-conditions}

Now we write  Eqs.\eqref{GR+T effects EOM} as:
\begin{equation}\label{eq:fR_MG}
    G_{\mu\nu}=\kappa\left(\mathcal{T}_{\mu\nu}+\mathcal{T}_{\mu\nu}^{geom}\right)=\kappa \tilde{\mathcal{T}}_{\mu\nu}\,.
\end{equation}
In this context, we use the notation $G_{\mu\nu}:=R_{\mu\nu}-\frac{1}2g_{\mu\nu}R$ to describe Einstein tensor, incorporating  correction  from  $f({\mathcal {Q, T}})$ gravity similar to those  discussed in $f(R)$ gravity \cite{DeFelice:2010aj, Capozziello:2011et} as:
\begin{equation}
    \mathcal{T}_{\mu\nu}^{geom}=\frac{1}{\kappa f_\mathcal{Q}}\left[\frac{1}{2}g_{\mu\nu}\left(f(\mathcal{Q})-f_{\mathcal{Q}}
\left(\mathcal{Q}\right)\mathcal{Q}\right)-2f_{\mathcal{Q} \mathcal{Q}}\left(\mathcal{Q}\right)P_{\phantom{\alpha}\mu\nu}^{\alpha}\partial_{\alpha}\mathcal{Q}-f_\mathcal{T}\left(\mathcal{T}_{\mu \nu}+\Theta_{\mu \nu}\right) \right] \equiv { \mathcal{T}}^{(\text{eff})}_{\mu\nu}\,.
\end{equation}
Then $\tilde{\mathcal{T}}_{\mu}{^\nu}$ can be written as $\tilde{\mathcal{T}}_{\mu}{^\nu}=diag(-\tilde{\rho} c^2, \tilde{p}_r, \tilde{p}_t, \tilde{p}_t)$.
\begin{figure*}
\centering
\subfigure[~NEC \& WEC ]{\label{fig:Cond1}\includegraphics[scale=0.27]{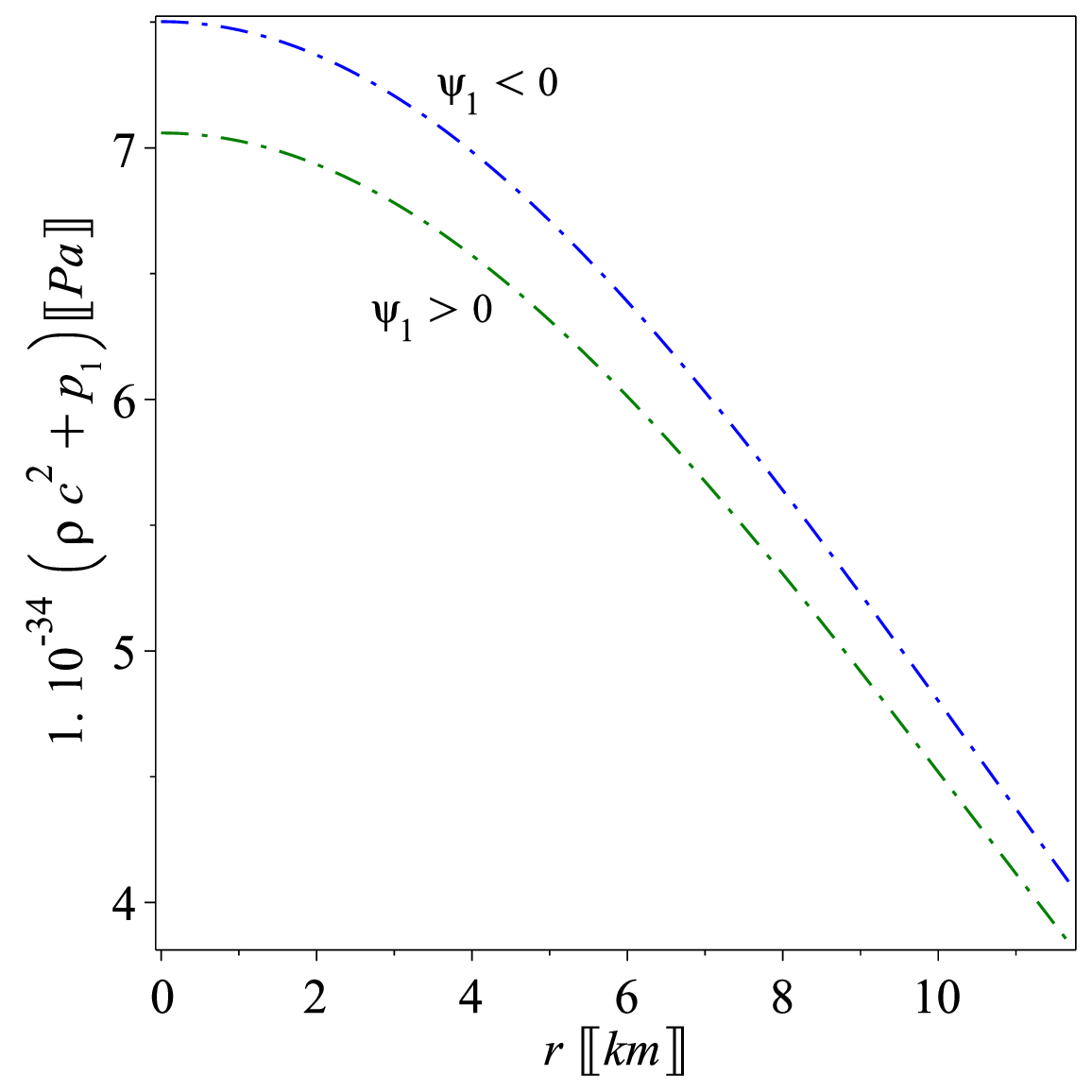}}\hspace{0.2cm}
\subfigure[~NEC \& WEC ]{\label{fig:Cond1}\includegraphics[scale=0.27]{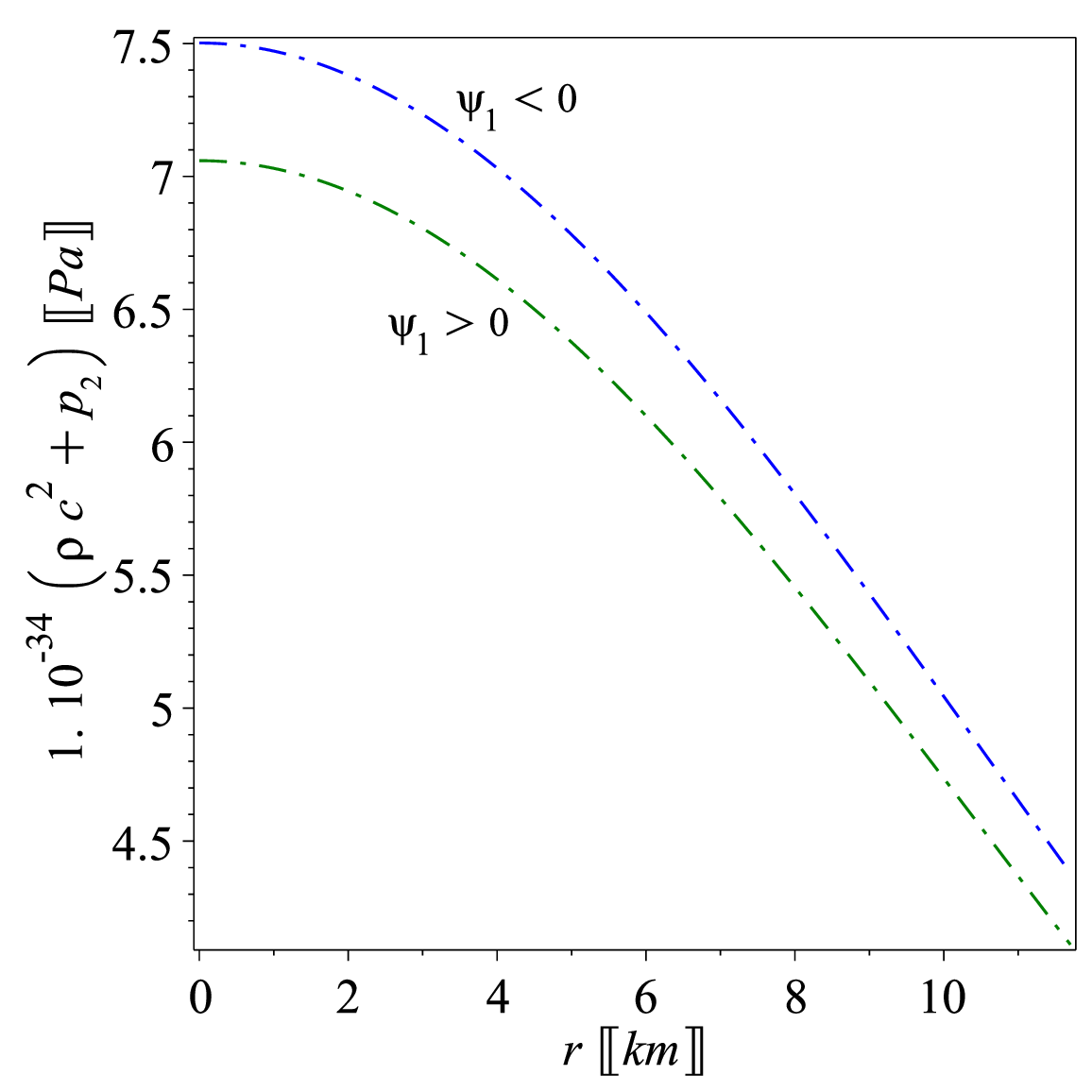}}\hspace{0.2cm}
\subfigure[~The SEC]{\label{fig:Cond3}\includegraphics[scale=.27]{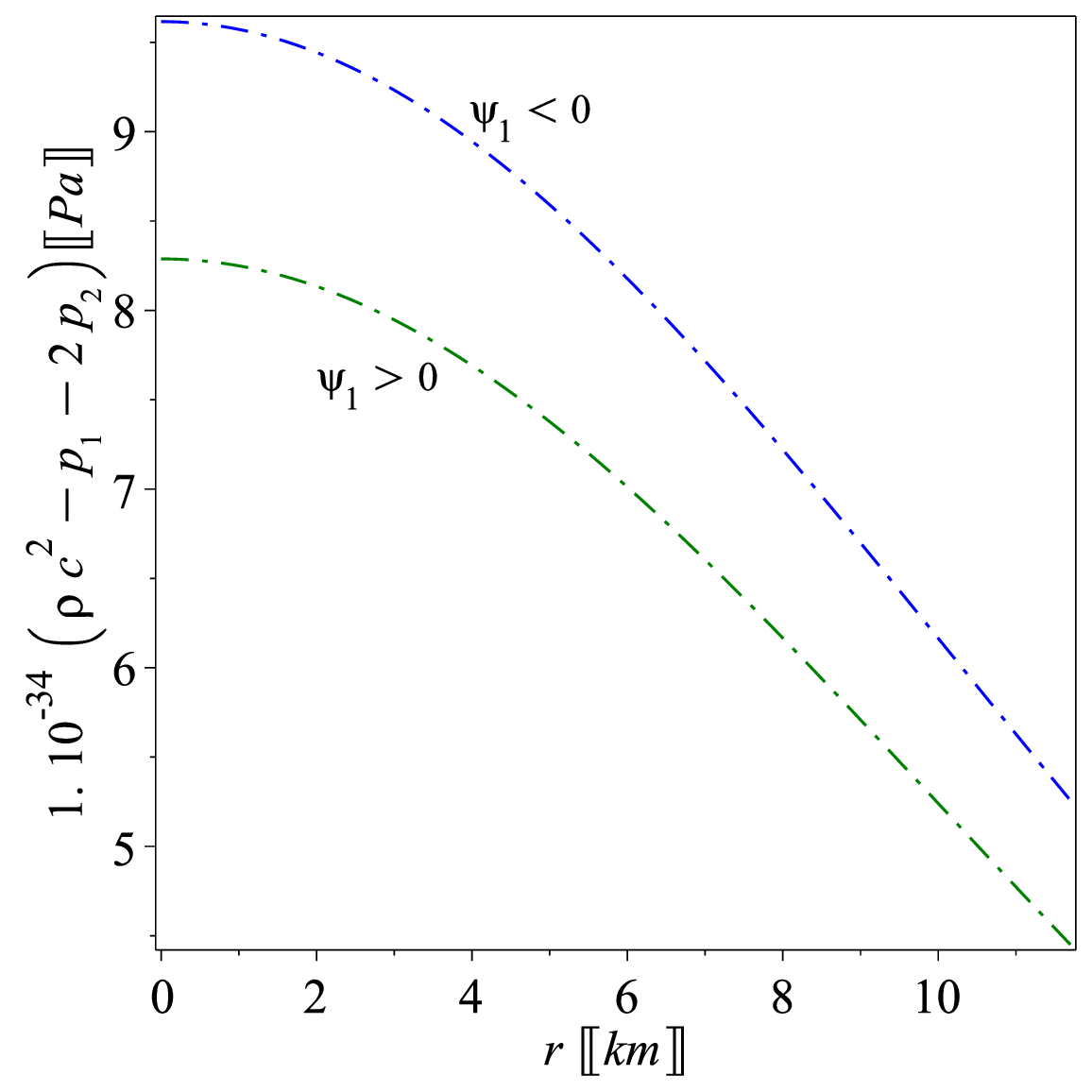}}\\
\subfigure[~DEC  in $r$-direction)]{\label{fig:Cond2}\includegraphics[scale=0.27]{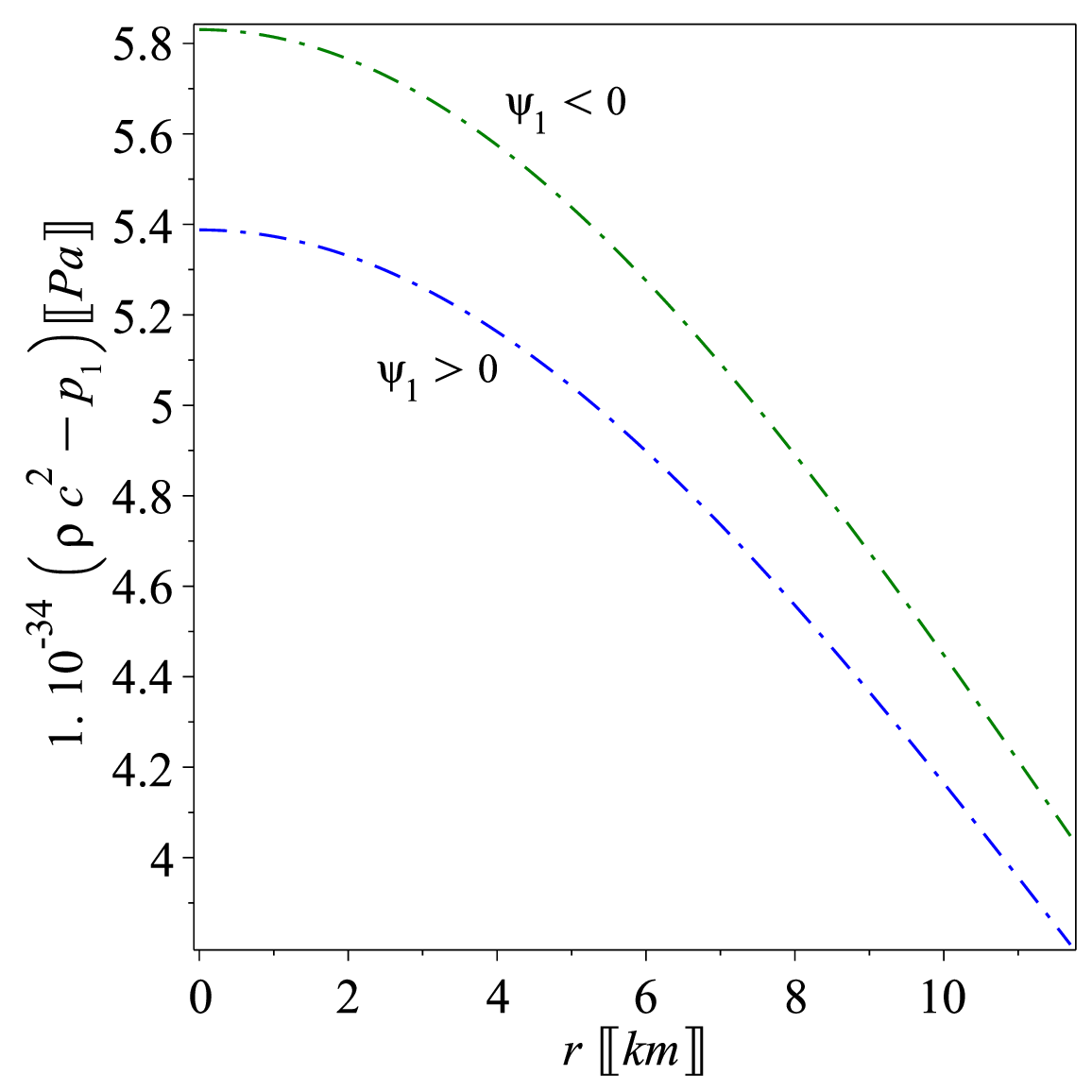}}\hspace{0.2cm}
\subfigure[~ DEC in tangential  direction]{\label{fig:DEC}\includegraphics[scale=.27]{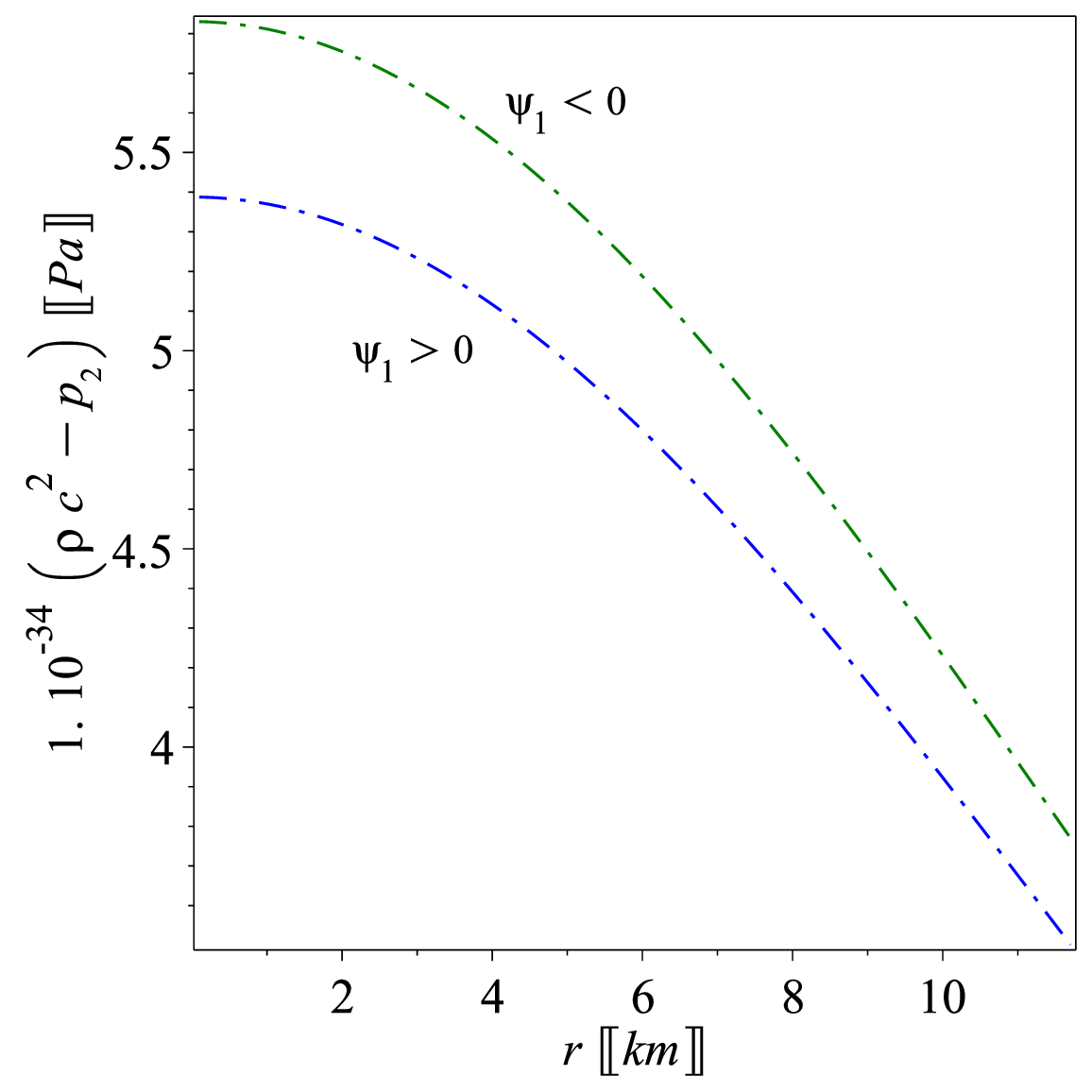}}\hspace{0.2cm}
%
\caption{The above plots demonstrate  that  $\bar{\mathcal{T}}_{\mu\nu}$ meets  the constraints of energy, as discussed in Subsection \ref{Sec:Energy-conditions}, for  pulsar J1748.9-2021. The scenarios with $\psi>0$ and $\psi<0$ related  to $\psi_1=0.04$ and $\psi_1=-0.04$, respectively.}
\label{Fig:EC}
\end{figure*}

By applying  Raychaudhuri equation  it implies that: \textit{${R}_{\mu\nu} u^{\mu} u^{\nu} \geq 0$} and \textit{${ R}_{\mu\nu} n^{\mu} n^{\nu} \geq 0$}.  It is worth noticing that \textit{${ R}_{\mu\nu}$},  can be written as
${ R}_{\mu\nu}=\kappa\left(\tilde{\mathcal{T}}_{\mu\nu}-\frac{1}{2} g_{\mu\nu} \tilde{\mathcal{T}}\right)$. By keeping this in mind, the energy conditions can be extended to
  $f({\cal Q})$ gravity as  \cite{Capozziello:2014bqa}:
\begin{itemize}
  \item[1.]  $ \tilde{\rho} c^2+ \tilde{p}_1 > 0$, $\tilde{\rho} c^2+\tilde{p}_2 > 0$, and $\tilde{\rho}\geq 0$.
  \item[2.] $\tilde{\rho} c^2+  \tilde{p}_2 \geq 0$, and  $\tilde{\rho} c^2+ \tilde{p}_1 \geq 0$.
  \item[3.]  $\tilde{\rho} c^2+\tilde{p}_1 \geq 0$,  $\tilde{\rho} c^2+\tilde{p}_2 \geq 0$, and $\tilde{\rho} c^2+\tilde{p}_1+2\tilde{p}_2 \geq 0$.
  \item[4.]  $\tilde{\rho} c^2-|\tilde{p}_1| \geq 0$, $\tilde{\rho} c^2-|\tilde{p}_2| \geq 0$,and $\tilde{\rho}\geq 0$.\\
\end{itemize}
Figures \ref{Fig:EC}  ensure that the present solution of pulsar {\textit SAX J1748.9-2021} meets all  requirements.
\subsection{Conditions of causality }\label{Sec:causality}

\begin{figure*}
\centering
\subfigure[~Speed of sound in the radial direction]{\label{fig:vr}\includegraphics[scale=0.28]{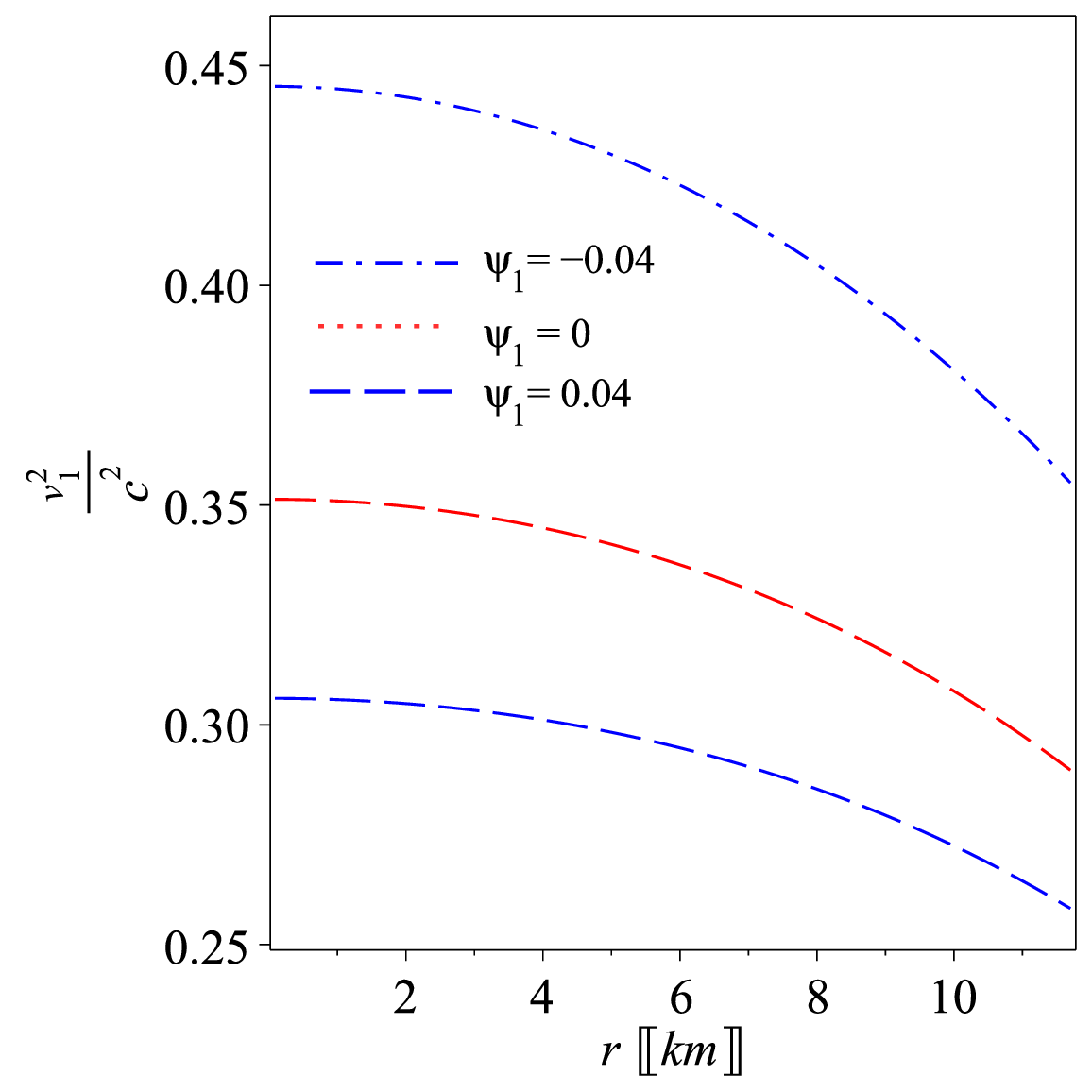}}\hspace{0.2cm}
\subfigure[~Speed of sound in the tangential  direction]{\label{fig:vt}\includegraphics[scale=.28]{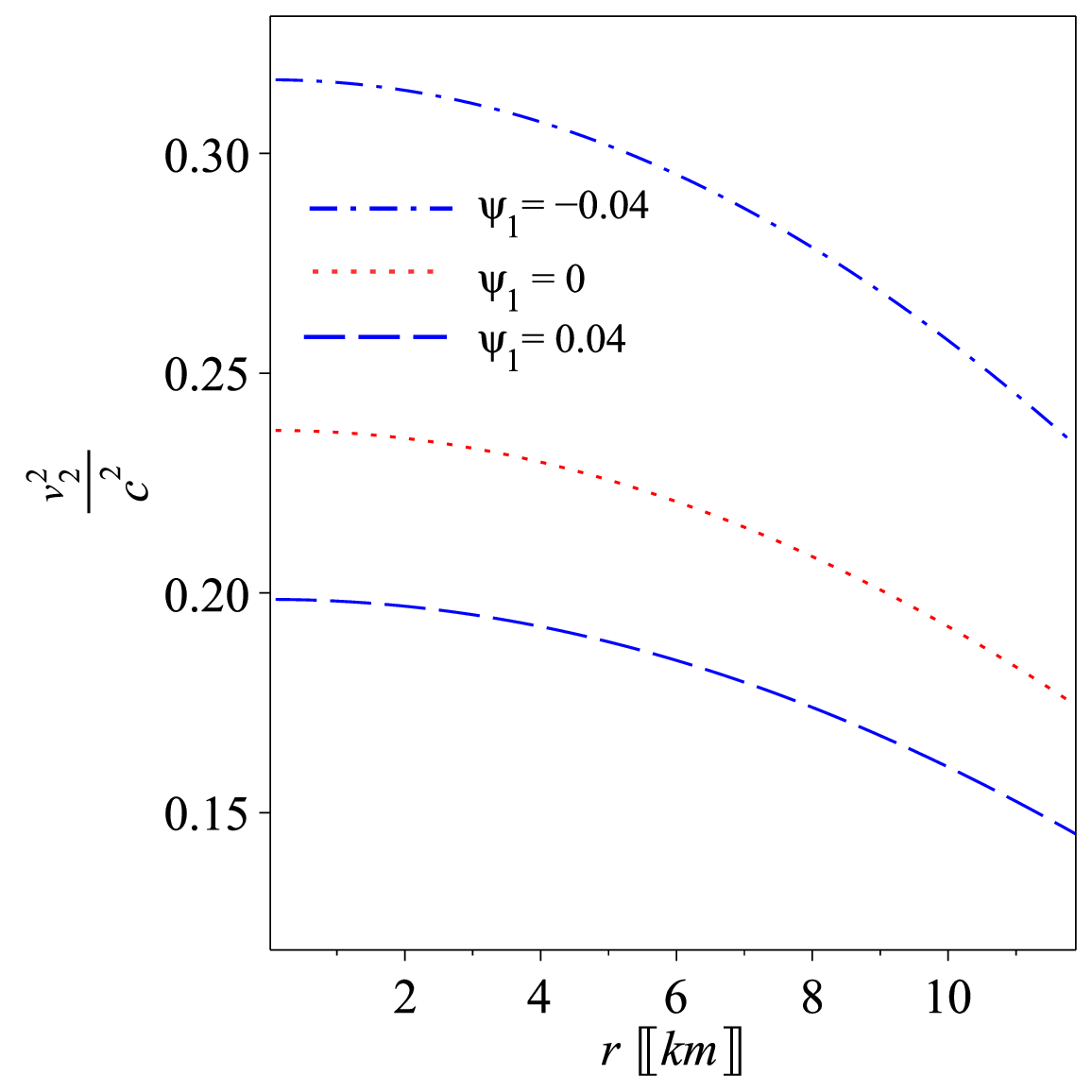}}\hspace{0.2cm}
\subfigure[~Stability in the presence of strong anisotropy]{\label{fig:vt-vr}\includegraphics[scale=.28]{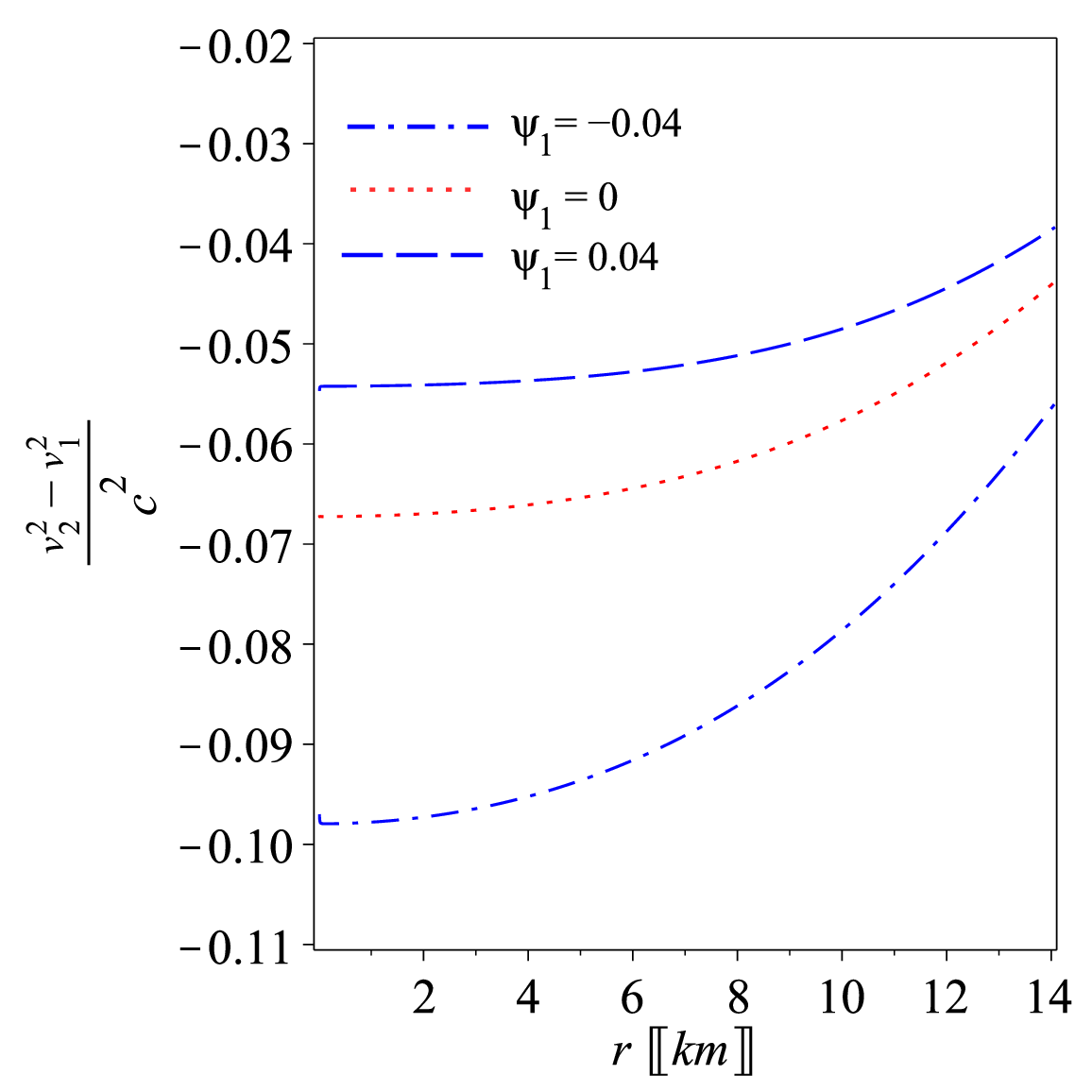}}
\caption{The sound speed  within  {\textit SAX J1748.9-2021} for $\psi_1=0,~ \pm 0.04$ is shown as: \subref{fig:vr} and \subref{fig:vt}  describe $v_1{}^2$ and $v_2{}^2$, respectively, as given in \eqref{eq:sound_speed}. \subref{fig:vt-vr} explains that the model complies with the stability condition $(v_2^2-v_1^2)/c^2 < 0$.}
\label{Fig:Stability}
\end{figure*}

One of the key factors defining physical structures is the causality condition. Now we  can be defined $v_1$ and $v_2$ as:
\begin{equation}\label{eq:sound_speed}
  v_1^2 =  \frac{ d{ p}_1}{d { \rho}}=  \frac{p'_1}{{ \rho'}}, \quad
  v_2^2 = \frac{d{  p}_2}{d{   \rho}}= \frac{p'_2}{{ \rho'}}\,.
\end{equation}
The gradients of  density and pressure components can be determined as provided by Eqs. (\ref{19}). We express $v_1$ and $v_2$ within SAX J1748.9-2021 for different numerical values of   $\psi_1$, as investigated in Figs. \ref{Fig:Stability}\subref{fig:vr} and \subref{fig:vt}. Those figures investigate that $0\leq {v_1^2}/c^2\leq 1$ and $0\leq {v_2^2/c^2} \leq 1$, verifying the causality and stability constraints. Moreover,  Fig. \ref{Fig:Stability}\subref{fig:vt-vr}, shows $-1< (v_2^2-v_1^2)/c^2 < 0$ throughout  SAX J1748.9-2021  \cite{Herrera:1992lwz}.

It is worth noticing that the speed sound, in both  directions, exhibits variations as shown  in Figs. \ref{Fig:Stability}\subref{fig:vr} and \subref{fig:vt}.

\subsection{Equilibrium of hydrodynamic forces and  adiabatic indices}\label{Sec:TOV}
It is  known that, when the adiabatic index $\gamma$ of a specific equation of state (EoS) is higher than a certain value $4/3$, there is no maximum limit to the mass of a stable configuration, i.e.,  $\gamma > 4/3$. In contrast, a stable configuration in Newtonian gravity requires that $\gamma < 4/3$. However, it has been shown that a pulsar can maintain stability against radial disturbances even with $\gamma > 4/3$. In order to address this issue, we establish the adiabatic index \cite{Chandrasekhar:1964zz,chan1993dynamical} in the following manner:
\begin{equation}\label{eq:adiabatic}
{\gamma}=\frac{4}{3}\left(1+\frac{{ \sigma}}{r|{  p}'_1|}\right)_{max},\qquad \qquad
{\Gamma_1}=\frac{{\rho c^2}+{p_1}}{{p_1}}{v_1^2}, \qquad \qquad
{\Gamma_2}=\frac{{\rho c^2}+{p_2}}{{p_2}}{v_2^2 .}
\end{equation}
Clearly, when isotropy is present ($\sigma=0$), we get $\gamma=4/3$. In the case of mild anisotropy ($\sigma<0$), resembling Newtonian theory, we find $\gamma<4/3$, in agreement with the typical stability criterion. Nevertheless, for significant anisotropy ($\sigma>0$) like in this research, we observe $\gamma>4/3$. Equilibrium is reached when $\Gamma$ equals $\gamma$, and it necessitates that $\Gamma$ must be greater than $\gamma$ \cite{chan1993dynamical,1975A&A....38...51H}.

\begin{figure}
\centering
\subfigure[~$\gamma$]{\label{fig:gamar1}\includegraphics[scale=0.28]{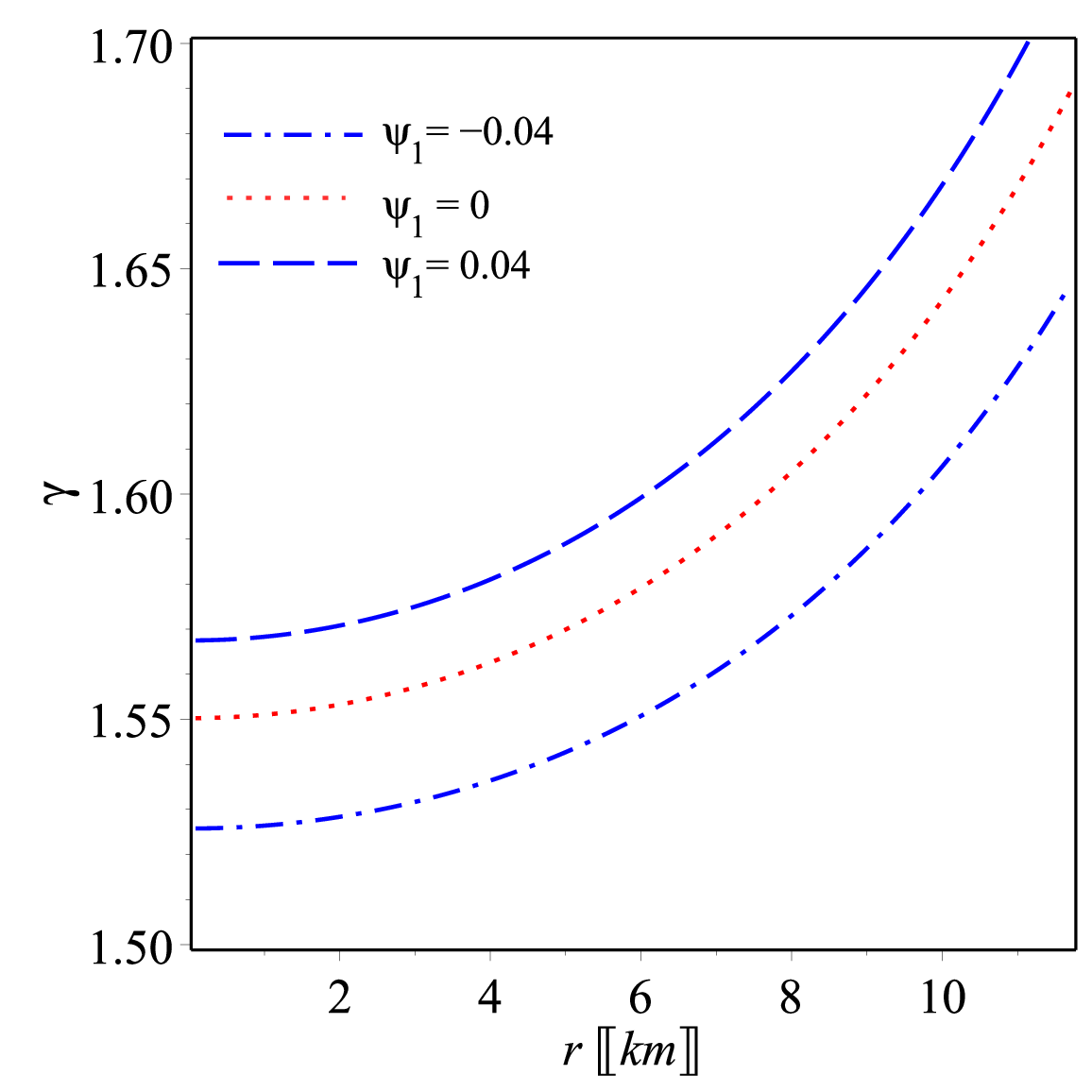}}\hspace{0.2cm}
\subfigure[~$\Gamma_1$]{\label{fig:gamar}\includegraphics[scale=0.28]{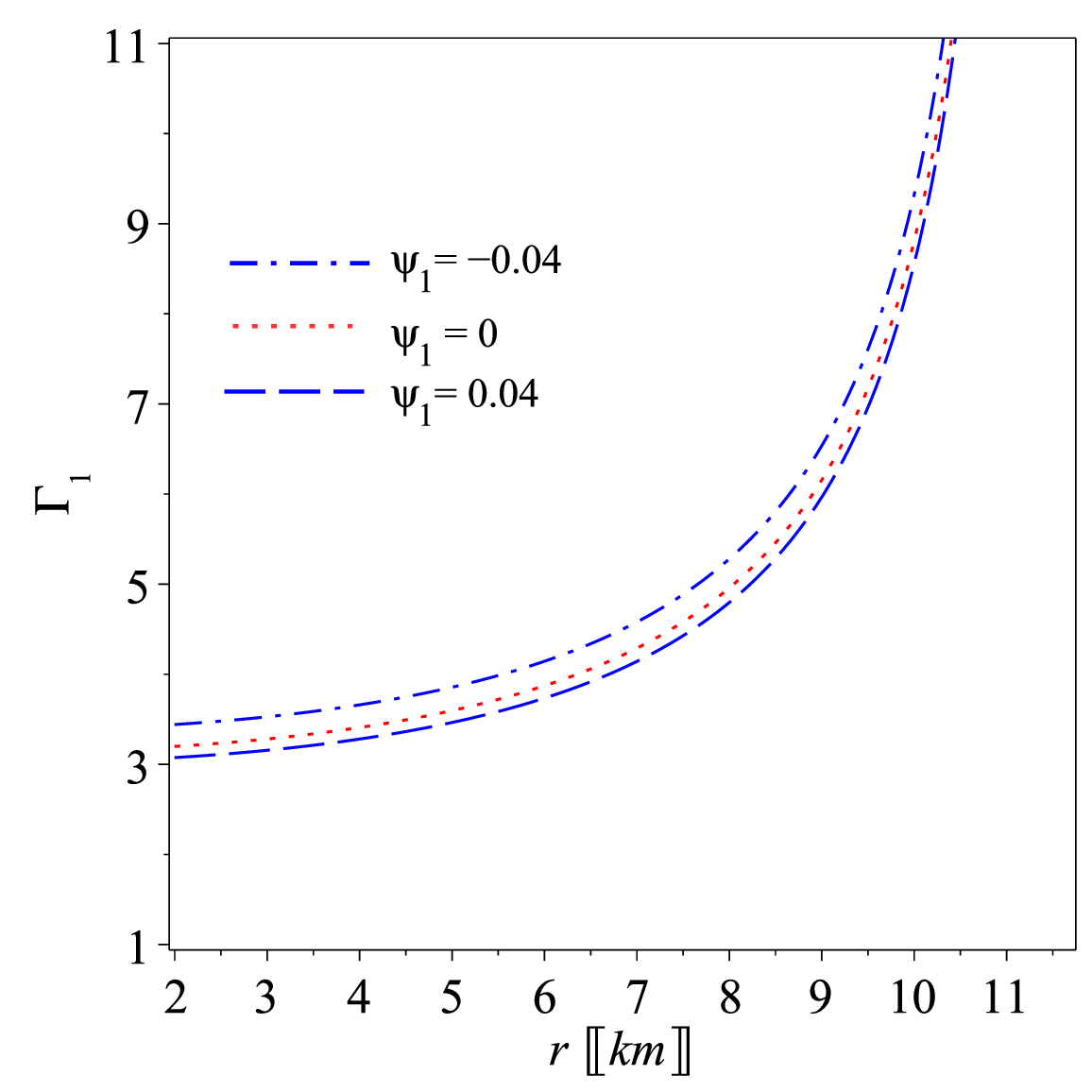}}\hspace{0.2cm}
\subfigure[~$\Gamma_2$]{\label{fig:gamat}\includegraphics[scale=0.28]{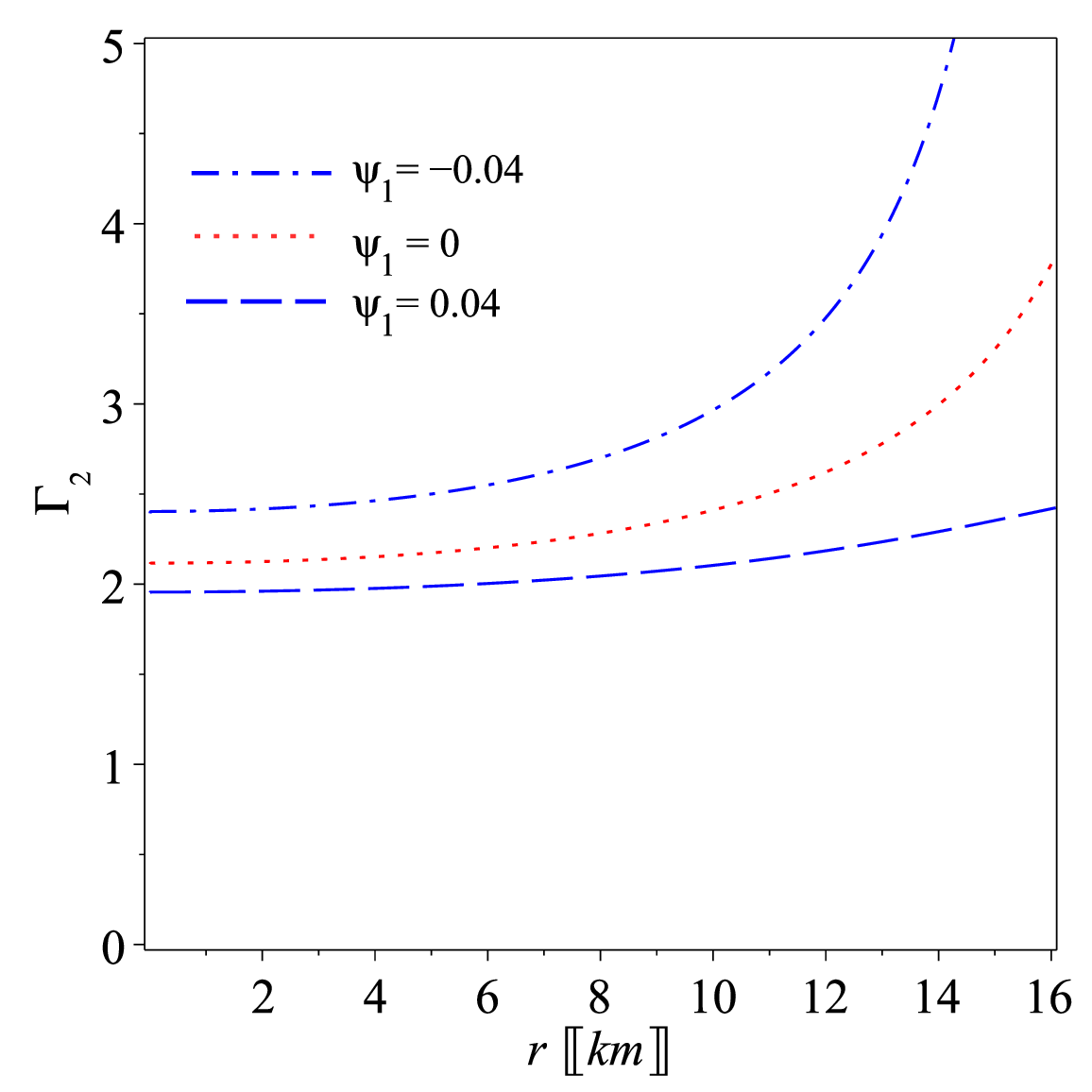}}
\subfigure[~Radial adiabatic index]{\label{fig:tov}\includegraphics[scale=0.30]{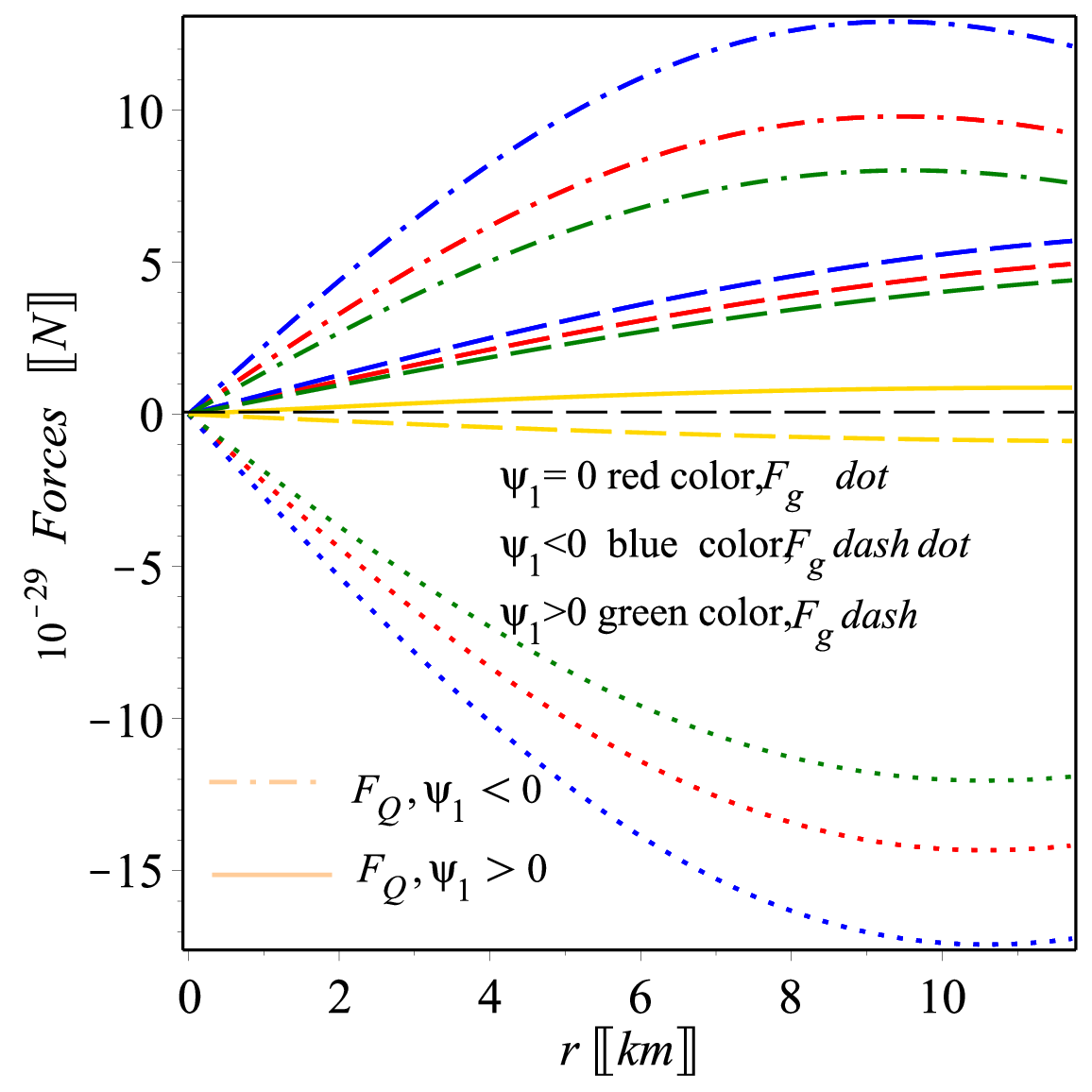}}
\caption{The indices of adiabatic, as provided  in \eqref{eq:adiabatic}, for  {\textit SAX J1748.9-2021}. These graphs show that the pulsar is stable, with $\gamma$ being greater than 4/3 and both $\Gamma_t$ and $\Gamma_r$ surpassing $\gamma$ within the pulsar's interior. Complying with the terms is crucial for a robust anisotropic fluid.}
\label{Fig:Adiab}
\end{figure}

Using the field equations and equations ({\color{blue} 4}), along with the gradients from Eq.~(\ref{eq:pr_grad}), we show that our $f({\cal Q,T})$ gravity creates a reliable anisotropic model for the pulsar SAX J1748.9-2021 with various numerical values of $\psi_1$, depicted in Figure \ref{Fig:Adiab}.

We will now analyze TOV equation in $f(\mathcal {Q,T})$ gravity in the following manner:

\begin{equation}\label{eq:RS_TOV}
{\mathcal F_a}+{\mathcal F_g}+{\mathcal F_h}+{\mathcal F_{\cal T}=0}\,.
\end{equation}
In addition to force $F_{\cal T}$, the theory considers the inclusion of gravitational, anisotropic, and hydrostatic forces known as ${\mathcal F_g}$, ${\mathcal F_a}$, and ${\mathcal F_h}$. These are described as:
\begin{eqnarray}\label{eq:Forces}
  {\mathcal F_a} =&\frac{ 2{\mathit  \Delta}}{\mathit r} ,\qquad
  {\mathcal F_g} = -\frac{{\mathit  M_g}}{r}({\mathit  \rho c^2}+{\mathit p_r})e^{\varepsilon/2} ,\qquad\nonumber\\
  {  \mathcal F_h} =&-{\mathit  p'_t} ,\qquad
  {\mathcal F_{\cal T}}  =\psi_1({  c^2 \rho}'-{ p}'_r-2{  p}'_{t})\,.
\end{eqnarray}
The force of gravity is indicated by $F_g$, where we have defined $\epsilon := \zeta - \zeta_1$, and $M_g$ stands for the gravitational mass of a closed system in the 3D space volume ${\mathit V}$ (at a fixed time $t$), which is described as \cite{1930PhRv...35..896T}:
\begin{eqnarray}\label{eq:grav_mass}
{\mathit M_g(r)}&=&{\int_{\mathit V}}\Big(\mathcal{{\tilde{\mathcal T}}}{^r}{_r}+\mathcal{\tilde{\mathcal T}}{^\theta}{_\theta}+\tilde{\mathcal{T}}{^\phi}{_\phi}-\tilde{\mathcal{T}}{^t}{_t}\Big)\sqrt{-g}\,dV\nonumber\\
&=&e^{-\zeta}(e^{\zeta/2})'  e^{\zeta_1/2} r =\frac{1}{2} r \zeta' e^{-\epsilon/2}\,.
\end{eqnarray}
As a result, the formula for gravitational force is given by ${\mathcal F_g} = -\frac{\varepsilon r}{R^2}({\mathit \rho c^2}+{ p_1})$. Using the field equations (\ref{eq.11}) and gradients, we can show that $f({\cal Q,T})$ gravity satisfies \eqref{eq:RS_TOV}, providing a reliable model for the pulsar {\textit SAX J1748.9-2021} over various $\psi_1$ values, including GR ($\psi_1=0$), shown in the figure \ref{Fig:Adiab}.

The condition of strong anisotropy, $\Delta>0$, produces a force in the opposite direction of the negative gravitational force. Expanding the size of the star is vital for increasing its mass while still keeping a stable structure. In Figure \ref{Fig:Adiab}\subref{fig:tov}, it is evident that the additional force caused by quadratic gravity can either help sustain gravitational collapse when $\epsilon$ is positive, or somewhat hinder gravitational collapse when $\epsilon$ is negative. The results for the pulsar {\textit SAX J1748.9-2021} in Subsection \ref{data} ($\psi_1=0.04, M=2.174 M_\odot, {R_s} =12.07~\text{km}, C=0.56$) and ($\psi_1=-0.03, M=1.953 M_\odot, {R_s} =12.88~\text{km}, C=0.49557$) confirm this analysis.

\section{Equation describing equilibrium and relationship between mass and radius}\label{Sec:EoS_MR}

The properties of matter within neutron star cores continue to perplex many astrophysicists, as these cores can have densities several times greater than nuclear saturation density, a level that cannot be replicated in laboratories on Earth. While the equation of state (EoS) for neutron star matter remains uncertain, observations of neutron star masses and radii can help constrain it or rule out certain possibilities. Consequently, astrophysical observations can refine the mass-radius plot  with a specific EoS.

\begin{figure}[th!]
\centering
\subfigure[~Radial EoS]{\label{fig:RfEoSp}\includegraphics[scale=0.45]{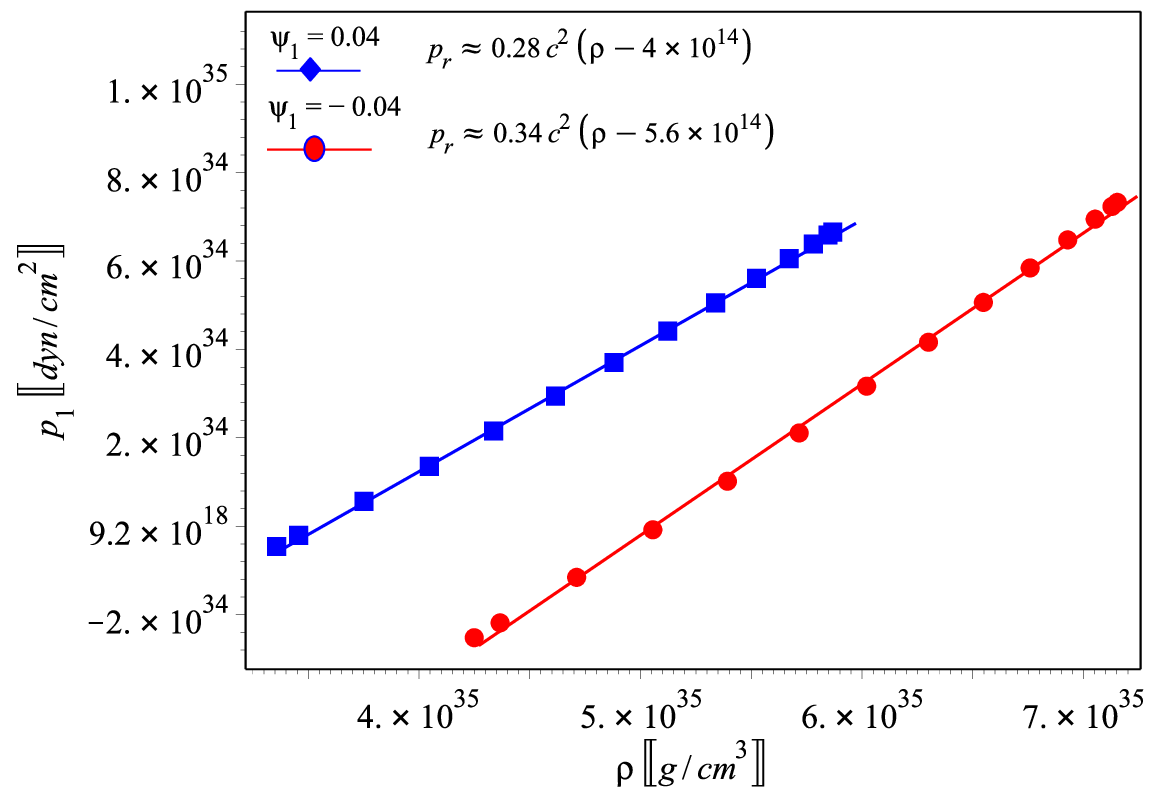}}
\subfigure[~Tangential EoS]{\label{fig:TEoSn}\includegraphics[scale=0.45]{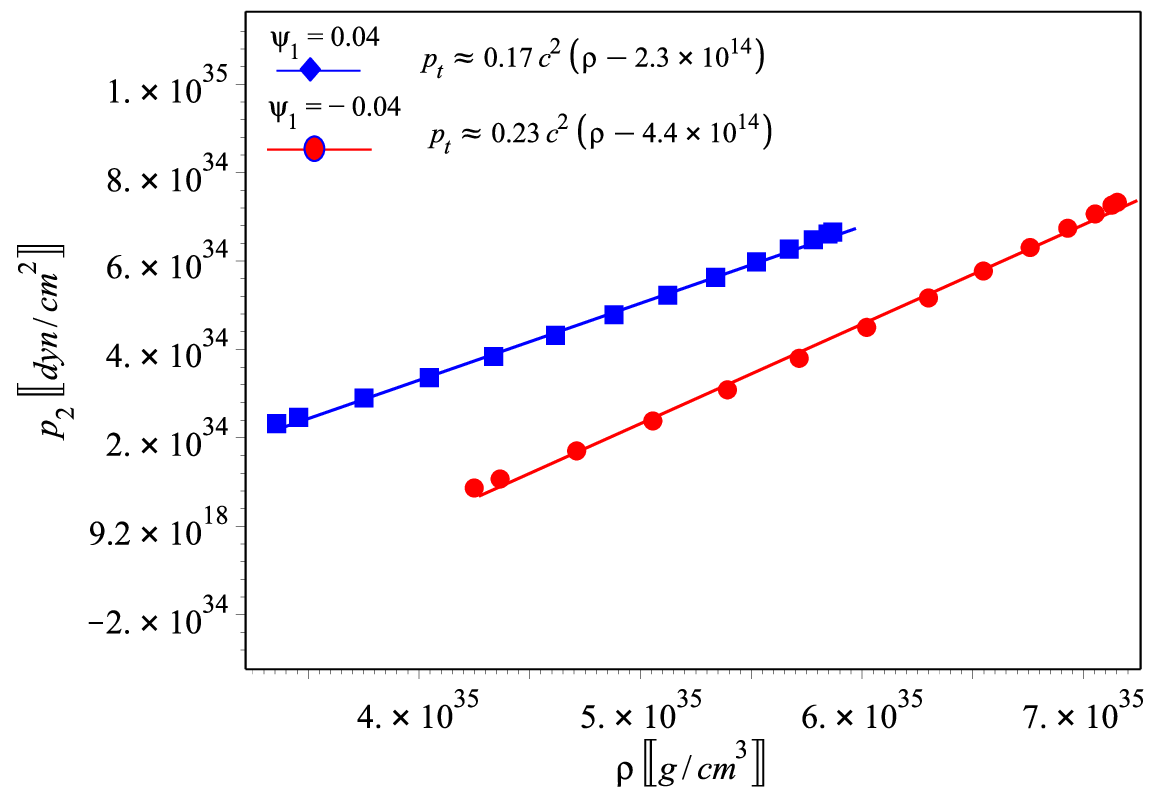}}
\caption{The best fit EoSs of the pulsar J0740+6620: \subref{fig:RfEoSp} We generate a sequence of data  for $\rho$ and $p_1$ by virtue of Eqs. \eqref{19} for $\psi_1=\pm 0.04$, the points fit well with linear EoS style. \subref{fig:TEoSn} Similarly for $p_t$ EoSs with $\psi=\pm 0.04$ the data fit well with linear model.
}
\label{Fig:EoS}
\end{figure}

 Using the numerical values given in \ref{data} for  {\textit SAX J1748.9-2021} and  Eqs. \eqref{19}, we create  the sequences shown in Fig. \ref{Fig:EoS}.

The data align well with a linear form for both scenarios: When \textit{$\psi_1=0.04$}, the fit equations  are \textit{$p_1 [dyn/cm^2]\approx 0.28 c^2(\rho-4 \times 10^{14}[g/cm^3])$}, \textit{ $p_2  [dyn/cm^2]\approx 0.17c^2(\rho-2.3\times 10^{14} [g/cm^3])$}. When  \textit{$\psi_1=-0.03$}, the  fit equations are \textit{$p_1 [dyn/cm^2]\approx 0.34 c^2(\rho-5.6 \times 10^{14}{[g/cm^3]})$},  \textit{$p_2 [dyn/cm^2]\approx 0.23c^2(\rho-4.4\times 10^{14}[g/cm^3])$}.

\begin{figure*}[t]
\centering
\subfigure[~Compactness ]{\label{fig:Comp}\includegraphics[scale=0.4]{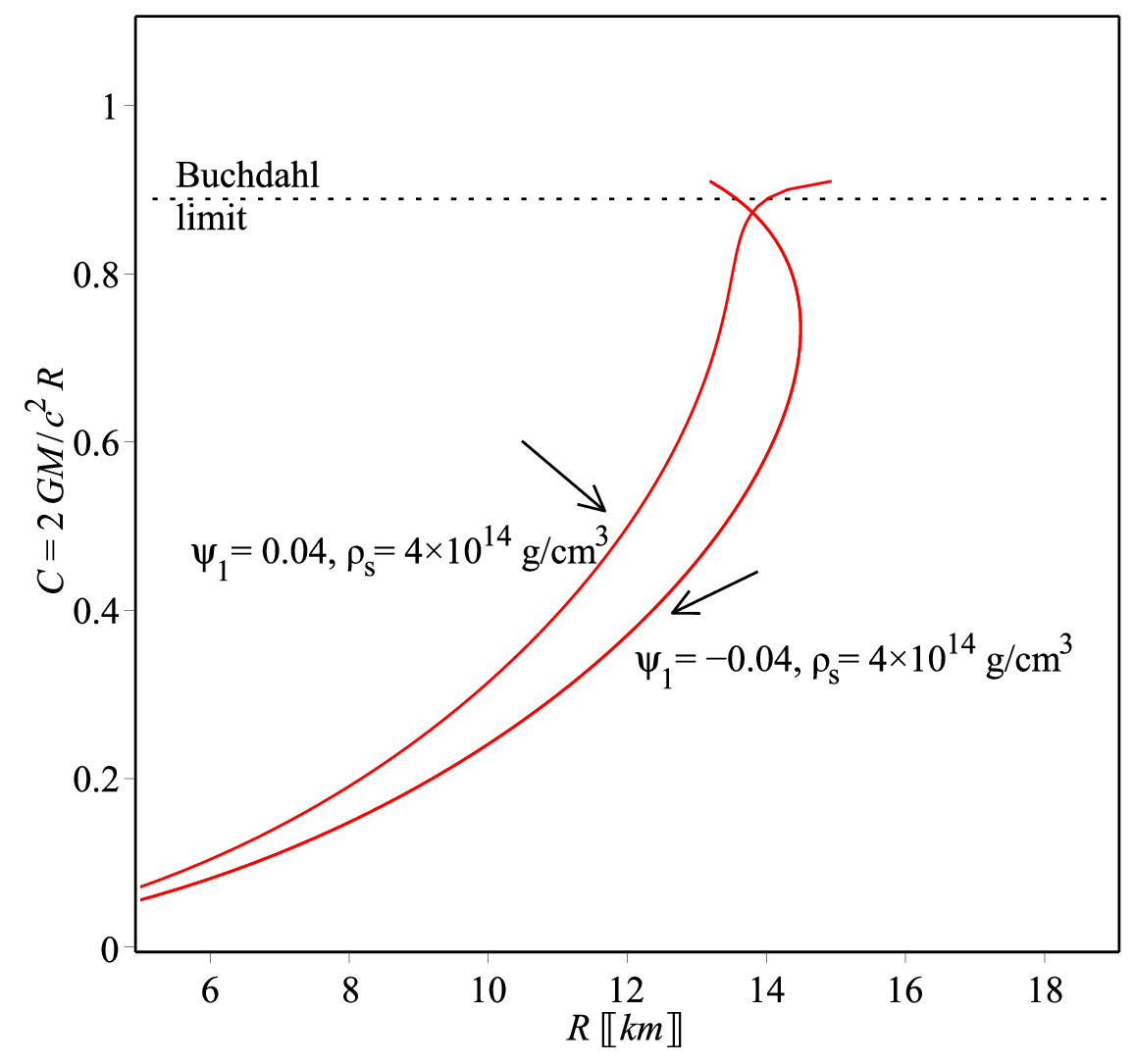}}\hspace{0.5cm}
\subfigure[~MR diagram]{\label{fig:MR}\includegraphics[scale=.4]{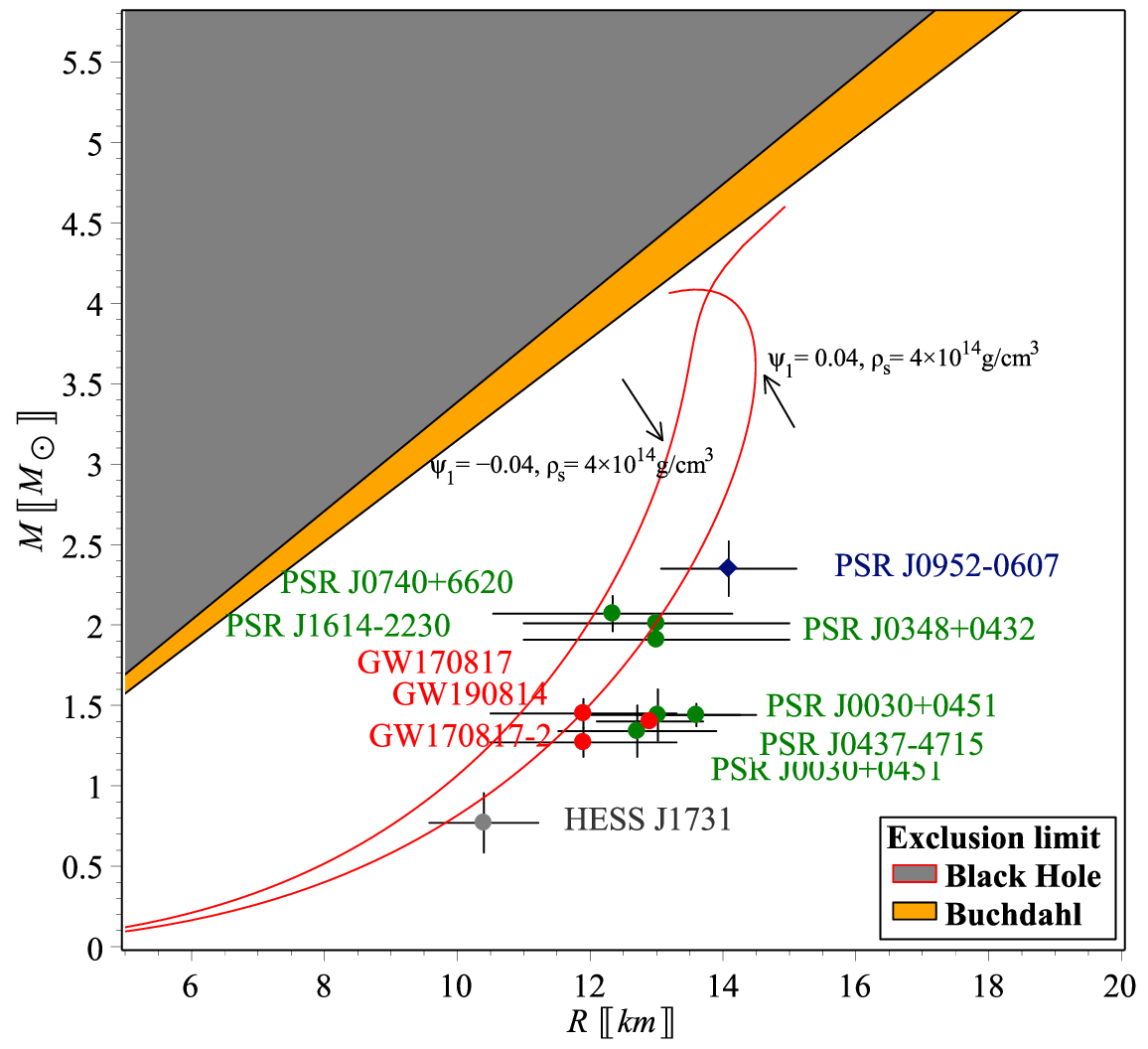}}
\caption{\subref{fig:Comp} The compactness-radius curves, using different values of $\psi_1$, which shows that the highest value of  $C= 0.9963$ for positive $\psi_1$ larger than  $C=8/9$ corresponding to the limit of Buchdahl.  and $C= 0.794$ for negative value of $\psi_1$ less than   $C=8/9$ (dotted horizontal line). \subref{fig:MR} The MR curves show  an upper mass of $M\approx 4.94 M_\odot$ with radius ${R_s}\approx 15.7$ km  for  positive $\epsilon_1$ and $M\approx 4.32 M_\odot$  with radius ${R_s}\approx 16.2$ km for negative $\epsilon_1$. 
\subref{fig:MR}  represent the Mass-radius relation}
\label{Fig:CompMR}
\end{figure*}

In this investigation of the pulsar J0740+662, we determine the Buchdahl limit: With $\psi_1=0.04$ we get $C\approx 0.94$, and with $\psi_1=-0.04$ it is $C\lesssim 0.93$, thus adjusting the GR Buchdahl limit slightly.

Because both situations only marginally alter the Buchdahl limit, we adhere to the usual restriction, denoted by $C\leq 8/9$, as illustrated   in Fig. \ref{Fig:CompMR}\subref{fig:Comp}. Initially, we consider the pairs ($\psi_1=0.03$, $\rho_\text{s}=4\times 10^{14}$ g/cm$^3$) and ($\psi_1=-0.04$, $\rho_\text{s}=4.0\times 10^{14}$ g/cm$^3$) on the surface derived from the top-performing EoSs. For any compactness parameter value $0 \leq C \leq 1$, we calculate the radius $R$ by solving the density profile \eqref{19}.

In this part, the Mass-Radius plots are shown in Fig. \ref{Fig:CompMR}\subref{fig:MR} when $\psi_1$ based on the best fit EoS obtained earlier, where the gravitational mass is determined using   $M= \frac{c^2 R}{2 G} (1-e^{-\varepsilon_2})$. As a result, we adopt the surface density boundary of $\rho_s=4\times 10^{14}$ g/cm$^{3}$ with $\psi_1=0.04$, resulting in a peak mass $M=4.15 M_\odot$ at a distance $R=14.45$ km. With a surface density of $\rho_s=4.0\times 10^{14}$ g/cm$^{3}$ and $\psi_1=-0.04$, a mass $M=4.66 M_\odot$ is reached at a radius of $14.9$ km.
\newpage
\section{Final Remarks}\label{con}

The objective of this study is to investigate the significance of the function $f({ \mathcal{Q, T}})$ within the framework of the SAX J1808.9-3658 pulsar. Understanding $f({ \mathcal{Q, T}})$ is crucial as it plays a pivotal role in describing the physical properties and behavior of compact objects like neutron stars. By analyzing its impact on the hydrostatic equilibrium and structural stability of such objects, we aim to deepen our comprehension of the fundamental mechanisms governing astrophysical phenomena, particularly in extreme environments like pulsars.
\begin{itemize}
\item
Our methodology involved employing exact analytical techniques to study the behavior of compact objects, specifically focusing on the ${\textit SAX J1808.9-3658}$ pulsar.
 Assumptions were made regarding the nature of the compact object, such as assuming it to be an anisotropic perfect fluid and considering specific forms of $f({ \mathbb{Q, T}})= \mathbb{Q}+\psi\mathbb{T}$  to model gravitational effects accurately. Additionally, observational constraints and theoretical considerations were taken into account  for the pulsar ${\textit SAX J1808.9-3658}$. The system we derived consists of three non-linear differential equation in five unknowns. To close the system we impose the ansatz of KB and get the explicate form of $\rho$, $p_r$, $p_t$ and $\Delta$.
\item
The study on the application of $f({ \mathcal{Q,T}})$ gravity theories to the {\textit SAX J1808.4-3658} pulsar has yielded several key findings and advancements:\\
1- We have shown that the dimensional parameter $\psi$ can take positive/negative value.\\
2-We have investigated that for {\textit SAX J1808.4-3658} the mass can exceed that of the GR, i.e., $\psi=0$  for   $\psi<0$ and the inverse is true when $\psi>0$.\\
3-We have shown that  the energy constraints  are verified for   {\textit SAX J1808.4-3658} for both cases of $\psi$.\\
4- We have investigated that the model under consideration did not violate the  conjecture conformal upper limit $v_1^2/c2 = 1/3$.\\
5- We have shown that our model has extra force due to the coupling constant $\psi$. When $\psi<0$ or $\psi>0$ we have shown that the TOV equation is satisfied and give net force equal to zero. Moreover, we have shown that the adiabatic indexes are greater than 4/3. All these means that the model under consideration is stable. \\
6-By utilizing equations \eqref{19} for $\psi_1=\pm 0.04$, we have produced a series of points for the density, radial pressures, and tangential pressures. We have demonstrated that the data fix with the linear EoS style as illustrated in \subref{fig:TEoSn}.\\
7- Finally, we draw the mass radius relation for certain surface density, i.e., $\rho_s=4 \times 10^{14}\text{[g/cm$^3$]})$ and showed that we have maximal mass when $\psi<0$. \\
\item Overall, the  study's key findings contribute significantly to the ongoing exploration of alternative gravity theories and their implications for understanding the nature of compact objects like pulsars.  These contributions pave the way for further exploration and understanding of the fundamental forces governing the universe's most extreme phenomena.

\end{itemize}
It is important to note that we have achieved more stable anisotropic stellar structures with the $f(\mathcal{Q,T})$ terms compared to GR. Viable and stable compact stars are present in this revised theory, it can be inferred. This study focuses specifically on the linear scenario of $f(\mathbb{Q, T})$, which is $\mathbb{Q}+\psi\mathbb{T}$. Different formulas for $f({ \mathbb{Q, T}})$ will be examined in our upcoming research.
%

\end{document}